\documentclass[showpacs,aps,prb,floats,twocolumn]{revtex4}
\usepackage{amsmath}

\usepackage{bm,braket}
\usepackage{lipsum}

\usepackage{hyperref}
\usepackage{xcolor}

\hypersetup{colorlinks=false,linkbordercolor=red,linkcolor=green,pdfborderstyle={/S/U/W 1}}

\usepackage{float}
\usepackage{graphicx}
\usepackage{amsfonts}
\usepackage{amssymb}
\usepackage{color}
\usepackage{ulem}
\newcommand {\be}{\begin{equation}}
\newcommand {\ee}{\end{equation}}
\newcommand {\bea}{\begin{eqnarray}}
\newcommand {\eea}{\end{eqnarray}}
\newcommand {\nn}{\nonumber}
\renewcommand{\v}[1]{\ensuremath{\mathbf{#1}}} 

\newcommand{\abs}[1]{\left| #1 \right|} 
\newcommand{\avg}[1]{\left< #1 \right>} 

\def\sm{\sigma^-}
\def\smdag{\sigma^+}
\def\f{\textbf{f}}
\def\fdag{\textbf{f}^\dagger}

\def\r1{\textbf{r}}
\def\R{\textbf{R}}
\def\a{{a}}
\def\adag{{a}^\dagger}

\begin{document}

\abovedisplayskip=7pt
\abovedisplayshortskip=7pt
\belowdisplayskip=7pt
\belowdisplayshortskip=7pt

\title{Quantum theory of light emission from quantum dots coupled to structured photonic reservoirs and acoustic phonons}

\author{Kaushik Roy-Choudhury$^*$ and Stephen Hughes}
\affiliation{Department of Physics, Queen's University, Kingston, Ontario, Canada, K7L 3N6}
\email{kroy@physics.queenu.ca}

\begin{abstract} 
Electron-phonon coupling in semiconductor quantum dots plays a significant role in determining the optical properties of excited excitons, especially the spectral nature of emitted photons. This paper presents a comprehensive theory and analysis of emission spectra from artificial atoms or  quantum dots coupled to structured photon reservoirs and acoustic phonons,
when excited with  incoherent pump fields. As specific examples of structured reservoirs, we chose a Lorentzian cavity and a coupled cavity waveguide, which are of current experimental interest. For the case of optical cavities, we directly compare and contrast the spectra from three distinct theoretical approaches to treat electron-phonon coupling, including  a Markovian polaron master equation, a non-Markovian phonon correlation expansion technique and a semiclassical linear susceptibility approach, and we point out the limitations of these models. 
For the cavity-QED polaron master equation, we give closed form analytical solutions to the phonon-assisted scattering rates in the weak excitation approximation, fully accounting for temperature,
cavity-exciton detuning and cavity dot coupling.
 We show explicitly why the semiclassical linear susceptibility approach fails to correctly account for phonon-mediated cavity feeding.  For weakly coupled cavities, we calculate the optical spectra using a  more general reservoir approach and explain its differences from the above approaches in the low Q limit of a Lorentzian cavity. We subsequently use this general reservoir  to calculate the emission spectra from quantum dots coupled to slow-light photonic crystal waveguides, which demonstrate a number of striking photon-phonon coupling effects. Our quantum  theory can be   applied to a wide range of photonic structures including photonic molecules and coupled-cavity waveguide systems.
\end{abstract}



\maketitle


\section{Introduction}\label{sec1}
Artificial atoms or quantum dots (QD) are  excellent candidates for solid-state quantum bits (qubits) and show promise for enabling scalable quantum information
processing  at optical frequencies~\cite{Yoshie,Bose1}. Embedding
QDs in
photon cavity structures  can 
manipulate the radiative decay rate by tailoring the local photon
density of states (LDOS). For example, one can enhance the 
spontaneous emission (SE) rate through the
 Purcell effect~\cite{Purcell},  or suppress  SE  leading to long lifetimes of around tens of ns~\cite{Dirk}, which is promising for low error rate quantum logic operations~\cite{Kim}. However being part of a solid-state lattice structure,  QD excitons are intrinsically coupled to the underlying phonon bath~\cite{Axt2}, which significantly reduces their coherence time on short time scales. Phonon dressing of QD emission manifests itself in a number of experimental observations, such as phonon-assisted 
inversion~\cite{Tom:PRL05,Hughes:PRL11,Glassl:PRB11, QuilterPRL15, Bounouar, Ardelt,PhysicsInversion},   damping and frequency shifts of driven Rabi oscillations~\cite{Forstner, Ramsay, Leonard} and excitation induced dephasing of Mollow triplet sidebands~\cite{Ulrich,Edo1}, which distinguishes QDs from simple two-level atoms~\cite{Weiler}. 
Over the past decade, several theories have been developed to address this issue. They range from the independent Boson model (IBM)~\cite{Axt2, Besombes}, correlation expansion techniques\cite{Forstner}, perturbative master equations (MEs)~\cite{Nazir1}, polaron MEs~\cite{Imamoglu,Roy2,Hohenester, Kaer}, variational MEs~\cite{Nazir3} and path integral calculations~\cite{Axt}.

\begin{figure}[t]
\includegraphics[width=0.88\columnwidth]{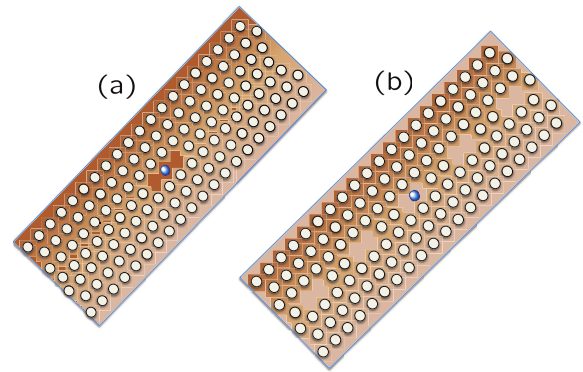}
\vspace{-0.cm}
\caption{\label{fig1}(Color online). \footnotesize{  
Two selected schematics showing a QD in a structured photonic reservoir. Photonic microcavity (a) and a coupled-cavity
waveguide (b) using a photonic crystal platform, containing an artificial atom or a coupled semiconductor QD.}}
\label{fig1}
\end{figure}

Coupling QDs to structured photonic reservoirs like photonic crystals
can be used for  building a scalable quantum light-matter interface~\cite{Kim} and  allows quantum control of light-matter interactions~\cite{Bose1}.
One common description for the QD-medium light  interactions is to treat the QD like a simple atom with discrete energy levels. 
However, the  phonon interactions may strongly influence the light-matter coupling and in turn modify the emission properties of the QD in a non-trivial manner~\cite{Ota, Milde, Arka,calicPRL}. For example, the well known Fermi's golden rule breaks down in weakly coupled photonic reservoirs which have photon correlations times that are comparable the the phonon bath and this results in enhanced or even suppressed SE~\cite{Kaushik}.
Experimentally,  phonon interactions  are clearly visible in the photon emission spectra, especially when  coupled to a high Q cavity, where Q is the quality factor. When electron-phonon coupling is included in the models,
polaronic approaches~\cite{Roy2} and linear susceptibility theories~\cite{HughesSC, savonaPRB} (where the IBM model is included in the QD polarizability) predict asymmetric vacuum Rabi splitting \cite{Milde, HughesSC, Jake} and strong off-resonant cavity feeding~\cite{Arka,HughesSC,savonaPRB}, which have  been observed experimentally~\cite{Ota, Arka,calicPRL}.

In this paper, we present a comprehensive analysis of emission spectra from QDs coupled to structured photonic reservoirs, in the presence of electron-phonons scattering. The development of a  general light-matter interaction theory for a QD exciton coupled to an arbitrary photon and phonon bath is extremely challenging and several approximate treatments have been employed to treat several regimes of interest. For example, the cavity-QED (cQED) polaron ME approach, which treats the cavity mode operator at the level of a system operator, is suitable for modelling high-Q cavities~\cite{Roy2}. If a 
weak-to-intermediate coupling regime with the photon bath is assumed, the polaronic ME theory can be extended to handle arbitrary structured {\it photon}
reservoirs using a photon bath approximation~\cite{Kaushik}.
A semiclassical approach has also been used to model the spectrum in the linear excitation regime\cite{savonaPRB,HughesSC}, though is restricted in its range of excitation conditions and modelling true {\it quantum} light-matter interactions; we show explicitly the failure of such an approach for modelling cavity feeding over large QD-cavity detunings. Without cavity and photon bath coupling, a common perturbative approach to describing electron-phonon coupling has been through the correlation expansion technique, which has been used to explain phonon-induced damping of Rabi oscillations where electron-phonon coupling is treated   to second-order~\cite{Forstner, Forstner1}; we extend such an approach to study QDs in a simple cavity and make a direct comparison with the polaron ME approaches and the linear susceptibility model.

 Each of these approximate theories has its own merits and limitations, though there
 has been little investigation into how each of these approaches compare with each other, and, to the best of our knowledge, there has been no  quantum approach for treating the emitted spectrum for excited QDs in the presence of phonons and general photon reservoirs.
For the photon medium, we consider  examples of a simple photonic crystal cavity~\cite{Yoshie} (Fig.~\ref{fig1}(a)) and a coupled resonator optical waveguide (CROW)~\cite{Notomi1} (Fig.~\ref{fig1}(b)) which are relevant in the context of current semiconductor QD experiments. For the case of a high-Q cavity, we examine the case of  phonon-dressed vacuum Rabi splitting~\cite{Ota,Milde}, when strongly coupled QD-cavity are at resonance. For the far off-resonant case between the QD exciton and a Lorentzian cavity, we consider the scenario of phonon-assisted cavity feeding~\cite{Hohenester,HughesSC,Arka, savonaPRB,calicPRL}.
For the photonic cavity, the polaron-transformed QD-cavity~\cite{Imamoglu} interaction is simplified using a Born-Markov approximation~\cite{Roy2, Roy3}. This approximation is expected to be valid if the system dynamics is longer than the correlation time of the phonon bath. To investigate the possible breakdown of this approach, we utilize the non-Markovian correlation expansion approach~\cite{Forstner} to include a cavity, and we use this method to help distinguish the regimes in which the polaron ME approach is valid, since the latter is considerably simpler to use and can easily include in other quantum processes such as multi-photon effects.  We also compare the above two approaches with the predictions of a semiclassical but non-Markovian linear susceptibility theory~\cite{HughesSC,savonaPRB}, where the IBM lineshape is incorporated within the frequency-dependent polarizability function of the QD exciton.
The IBM lineshape shows  acoustic phonon sidebands that surround
the zero phonon line (ZPL).
We  show the clear breakdown of this approach and introduce a fully quantum approach
to treat the emission spectrum of a QD in the presence of any spectral shaped
photon reservoir coupling~\cite{Kaushik}, which is valid in the Purcell regime, i.e., 
for weak-to-intermediate coupling.
In the case of a simple leaky cavity,
  differences from the cQED polaronic ME and the reservoir polaron ME approaches arise, when the coupled cavity becomes increasingly lossy~\cite{Kaushik}.  Lastly, we use this general reservoir approach to investigate emission spectra of QDs from photonic crystal coupled cavity waveguides~\cite{Kaushik}. To the best of our knowledge, this is the first theory to study such spectral features from waveguides, without making a mean-field approximation for 
phonon coupling effects~\cite{Hoang}. In the mean-field limit, it has been common to assume that phonons reduce SE in a structured reservoir by a frequency independent constant $\avg{B}$~\cite{Hoang,Roy1}, where $\avg{B}$ is the the thermal average of phonon bath displacement operators $B^\pm$~\cite{Imamoglu} (also see Sec.~\ref{sec2a}) which  depends on the bath temperature. The polaronic reservoir
approach~\cite{Kaushik}, however, shows that SE can also be enhanced by phonons and depends on the local density of states (LDOS) of 
the photon bath reservoir; also note the 
  polaronic approaches include effects that are nonperturbative in phonon electron-coupling and can be used to model low to high temperatures of the phonon bath.

Our paper is organized as follows. In Sec.~\ref{sec2}, we introduce the key theoretical models. Sections~\ref{sec2a} and \ref{sec2b}  briefly review the polaron reservoir ME~\cite{Kaushik} and the polaron cQED ME~\cite{Roy2}, respectively.  Section~\ref{sec2c} presents  the correlation expansion approach~\cite{Forstner} with the coupling to a leaky cavity mode. In Sec.~\ref{sec2d}, we derive the incoherent emission spectrum when the QD is excited by a weak incoherent pump field and show how this is obtained from the  two-time correlation function; and we also present the emitted spectrum from an excited exciton  using linear susceptibility theory (Sec.~\ref{sec2d4}). In Sec.~\ref{sec2e} we discuss the phonon parameters used for calculations, and in Sec.~\ref{sec2f} we discuss the numerical complexity of the different approaches. In Section~\ref{sec3}, we present and discuss various numerical results. We begin with a discussion of emission spectra from a bare (i.e., with no photonic bath coupling) QD (Sec.~\ref{sec3a}) and thereafter present results on QDs coupled to structured photonic reservoirs.  Section~\ref{sec3b} investigates the scenario of phonon-dressed, resonant vacuum Rabi splitting (i.e., in the strong coupling regime) and reveals the limitations imposed by the Born-Markov approximation on existing polaron cQED approaches. Section~\ref{sec3c} treats the case of off-resonant cavity feeding under weak coupling conditions. The effects of accounting for intercorrelated photon and phonon bath dynamics is discussed for low Q cavities using the photonic reservoir theory~\cite{Kaushik} (Sec.~\ref{sec3c1}). The different predictions of the linear susceptibility theory in the context of off-resonant cavity feeding is also explained (Sec.~\ref{sec3c2}). Finally, in Sec.~\ref{sec3d}, we study in detail the emission spectrum from QDs weakly coupled to at photonic crystal waveguides; we use a model LDOS suitable for a slow-light CROW. Section~\ref{sec4} summarizes the main results and discusses future directions.
In Appendix~\ref{Appen1} we show the equivalence between the spectra obtained from a weak incoherent pump and an inverted exciton as the initial condition, since these two approaches are used with the MEs and linear susceptibility model, respectively.

\section{The Hamiltonian with photon bath and phonon bath coupling to a  Quantum dot exciton}\label{sec2}

\begin{figure}[t]
\includegraphics[width=0.99\columnwidth]{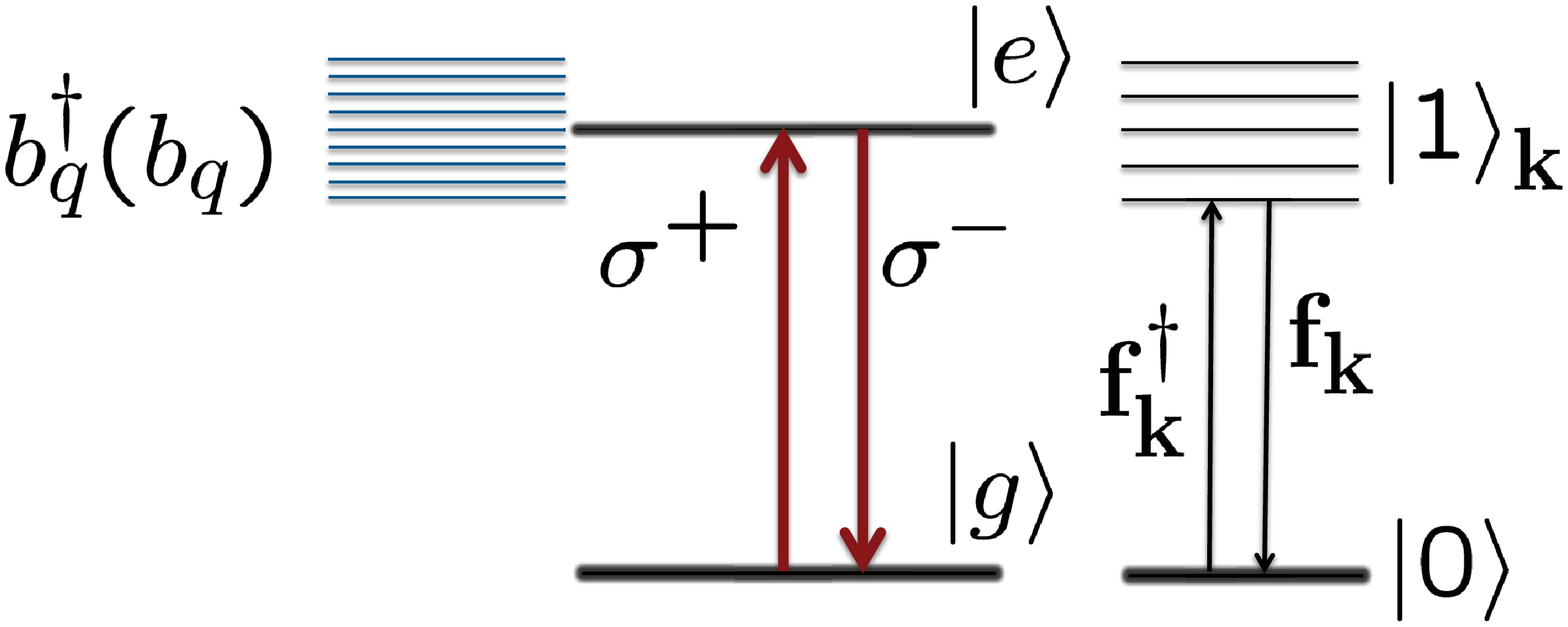}
\vspace{-0.cm}
\caption{(Color online). \footnotesize{ Energy level diagram of a neutral QD  (electron-hole pair) interacting with a phonon bath and a photon bath. The operator ${\bf f}^\dag_{\bf k}$
($b^{\dag}_q$) creates a photon (phonon).} }
\label{fig2}
\end{figure}

We consider a single neutral QD exciton (strong confinement limit) that is modelled as a two-level system coupled to a photon bath  and a phonon bath described by lowering operators $\f$ and $b_{q}$, respectively (see Fig.~\ref{fig2}).
Such a two-level approximation is valid for small epitaxial QDs (e.g., self assembled InGaAs/GaAs QDs), 
whose parameters are used for the current study.
Although real QDs have many exciton levels over a broad band of frequencies,
the coupling to one exciton has successfully explained a number of experiments
when probing single exciton dynamics, e.g., Refs.~\onlinecite{Weiler,Ramsay};
in addition, one can extend such an approach to add in other exciton levels~\cite{RongchunOL}.
 The deformation potential coupling with longitudinal acoustic (LA)  phonons play the strongest role in such QDs. The resultant phonon side-bands span a frequency range of $\approx \pm$ 5 meV around the $s$-shell transition which is much smaller than the energetic separation between $s$-shell and $p$-shell transitions ($\approx$ 25 meV)~\cite{Weiler}, thus validating a two-level approximation.  The lowering operator $\sm$ describes a transition between the QD states $\ket{e}$ and $\ket{g}$, separated in energy $\omega_x$. In a frame rotating at the frequency of the QD, the coupled system is described by the
following Hamiltonian~\cite{Kaushik}:
\begin{align}
\label{eq1}
    H &= \hbar\int d\v{r} \int_0^{\infty}d\omega\,\fdag(\r1,\omega,t)\f(\r1,\omega,t)  + \sum_q\hbar \omega_q b^{\dag}_qb_q \nn\\
&-\left[\smdag e^{i\omega_xt}\int_0^{\infty}d\omega\,\v{d}\cdot\v{E}^+(\r1_d,\omega,t) + \text{H.c.}\right]  \nn\\
 &+ \smdag\sm\sum_q\hbar\lambda_q(b^{\dag}_q+b_q),
\end{align}
where a dipole and rotating wave approximation is made to describe coupling between the QD and the photon bath. In (\ref{eq1}), a QD of dipole moment $\v{d}$ is assumed to be located at the spatial position $\r1_d$,  and the exciton-phonon coupling strength $\lambda_q$ is assumed to be real~\cite{Roy2}. The positive frequency component of the electric field operator $\v{E}^+(\r1_d,\omega,t)$   can be expressed in terms of the  electric field Green function $\v{G}(\r1,\r1';\omega)$ as $\v{E}^+(\r1,\omega,t) =i\int d\r1' \v{G}(\r1,\r1';\omega)\sqrt{\frac{\hbar}{\pi\epsilon_0}\epsilon_{\rm I}(\r1',\omega)}\f(\r1',\omega,t)$~\cite{Scheel};
in a medium  described by the dielectric constant $\epsilon({\bf r},\omega)=
\epsilon_{\rm R}({\bf r},\omega)+{\rm i}\epsilon_{\rm I}({\bf r},\omega)$,
${\bf G}({\bf r},{\bf r}')$ is the solution to Maxwell equations
at ${\bf r}$ to a point dipole oscillating at ${\bf r}'$ (without any QD coupling). This expression for the electric field operator is quite general and fully satisfies the Kramers Kr\"onig relations in a general photonic medium. In order to avoid any coherent pump induced dressing of QD emission (which is interesting but beyond the scope of the present paper), the QD is assumed to be weakly excited by an incoherent pump $P$~\cite{Carmichael}, which maps on to a range of typical experiments that are performed to measure the low pump  emitted spectrum for a QD-photonic structure.
In the linear regime, one can obtain exactly the same expression starting from an excited exciton, but this requires one to integrate over two times
instead on one; this equivalence is shown explicitly in Appendix~\ref{Appen1}.
 In a ME  approach, the incoherent pumping term is included as a Lindblad operator, $\frac{P}{2}L(\smdag)$, where $L(O) = 2O\rho O^{\dagger}-O^{\dagger}O\rho - \rho O^{\dagger}O$ and $P$ is the pump rate; the ME can also include other Lindblad operators to account for additional incoherent processes such as background spontaneous decay, $\frac{\gamma_0}{2}L(\sm)$, and QD pure dephasing, $\frac{\gamma_d}{2}L(\smdag\sm)$. Neglecting coupling to a structured photonic reservoir and phonons, these incoherent processes determine the line-shape of the QD emission spectrum, which is represented by a simple Lorentzian broadening of the ZPL. 

We stress that the above Hamiltonian is applicable for arbitrary photon and phonon baths and is very difficult to solve in  general. In the following subsections, the Hamiltonian is thus manipulated using different approximations  which includes photon coupling, phonon coupling, and decoherence processes in a self-consistent and physically meaningful way. We then use these approaches to  compute the linear spectrum in different QD photon reservoir coupling regimes, and we  explore the range of validity of the various models. 

\subsection{Photon-reservoir polaron master equation approach}\label{sec2a}
We begin with the Hamiltonian in (\ref{eq1}) and perform a polaron transformation. This unitary transform includes phonons to all orders but puts the Hamiltonian in an easier form to apply system-bath reservoir theory and perturbation theory, with polaron shifted interaction terms. Essentially we are employing a basis change using a unitary transform in which the bath-shifted exciton becomes the polaronic quasiparticle. In this way, we fully recover the IBM without photon interactions and can  include processes that give rise to the ZPL.

 The polaron transformation is given by $H' \rightarrow e^P He^{-P}$, where $P = \smdag\sm \sum_q\frac{\lambda_q}{\omega_q} (b^{\dag}_q-b_q)$~\cite{Imamoglu}, which yields a polaron-transformed Hamiltonian, 
\begin{align}
\label{eq2}
&H' =  \hbar\int d\v{r} \int_0^{\infty}d\omega\,\fdag(\r1,\omega,t)\f(\r1,\omega,t)+\sum_q\hbar \omega_qb^{\dag}_qb_q\nn,\\
&   -\left[B_+\smdag e^{i\omega'_xt}\int_0^{\infty}d\omega\,\v{d}\cdot\v{E}^+(\r1_d,\omega,t) + \text{H.c.}\right], 
\end{align}
where $B_{\pm} = \exp[\pm\sum_q\frac{\lambda_q}{\omega_q}(b_q-b^{\dag}_q)]$ are the coherent phonon bath displacement operators~\cite{Imamoglu}. The polaron shift $\Delta_P = \int_0^{\infty}d\omega\frac{J_{\rm pn}(\omega)}{\omega}$ also appears  in the polaron transform  and below we will assume this factor is absorbed in the polaron-shifted frequency of the QD, defined by $\omega'_x$ (= $\omega_x-\Delta_P$). In the continuum limit, the phonon coupling is determined by the phonon spectral function $J_{\rm pn}(\omega)$~\cite{Roy3}. If a weak interaction between the QD and the photon reservoir is assumed, a ME for the QD reduced density matrix $\rho$ can be derived by retaining terms up to second order in the polaron-shifted interaction Hamiltonian, $H'_{\text{I}} =-[B_+\smdag e^{i\omega'_xt}\int_0^{\infty}d\omega\,\v{d}\cdot\v{E}^+(\r1_d,\omega,t) + \text{H.c.}]$, which is also formally known as the Born approximation. The form of the time convolutionless~\cite{Breuer} (or time-local) interaction picture ME is then
\begin{align}
\label{eq3}
&\frac{\partial \tilde{\rho}(t)}{\partial t} =\nn\\
&-\frac{1}{\hbar^2}\int_0^t d\tau \text{Tr}_{\text{R}_{\text{ph}}} \text{Tr}_{\text{R}_{\text{pn}}}\{ [\tilde{H'}_{\text{I}}(t),[\tilde{H'}_{\text{I}}(t-\tau),\tilde{\rho}(t)\rho_{\text{R}}]]\},
\end{align}
where $\tilde{H'}_{\text{I}}(t) = \exp[iH'_{\text{R}}t/\hbar ]H'_{\text{I}}\exp[-iH'_{\text{R}}t/\hbar ]$ and $H'_{\text{R}}=\hbar\int d\v{r} \int_0^{\infty}d\omega\,\fdag(\r1,\omega,t)\f(\r1,\omega,t)+\sum_q\hbar \omega_qb^{\dag}_qb_q$. The trace operators $\text{Tr}_{\text{R}_{\text{ph}}}$ and $\text{Tr}_{\text{R}_{\text{pn}}}$ denote a trace over the photon and phonon reservoirs, which are assumed to be statistically independent with $\rho_{\text{R}} = 
\rho_{\text{R}_{\text{ph}}}\rho_{\text{R}_{\text{pn}}}$~\cite{Carmichael}.
 The trace over the photon reservoir~\cite{Tanas,DaraN} assumes thermal equilibrium and uses the relations, $\text{Tr}_{\text{R}_{\text{ph}}}[\f(\v{r},\omega),\fdag(\v{r'},\omega')] = [\tilde{n}(\omega)+1]\delta(\v{r}-\v{r'})\delta(\omega-\omega')$ and 
$\text{Tr}_{\text{R}_{\text{ph}}}[\fdag(\v{r},\omega),\f(\v{r'},\omega')] = \tilde{n}(\omega)\delta(\v{r}-\v{r'})\delta(\omega-\omega')$ where $\tilde{n}(\omega) \approx 0$, which is valid for optical frequencies.  The final form of the polaron reservoir ME in the Schr\"{o}dinger picture~\cite{Kaushik} and Markov limit ($t\rightarrow\infty$ in the integral of (\ref{eq3})) is
\begin{align}
\label{eq4}
\frac{d\rho}{d t} &= \frac{\tilde{\gamma}}{2}L(\sm)-i \Delta_{\rm Lamb}[\smdag\sm,\rho] +\frac{\gamma_0}{2}L(\sm)\nn\\
&+\frac{\gamma_d}{2}L(\smdag\sm)+\frac{P}{2}L(\smdag),
\end{align}
 where the SE rate of the QD into the structured reservoir~\cite{Kaushik, DaraN} is 
\begin{align}
\label{eq5}
\tilde{\gamma} = 2\int_0^{\infty}\text{Re}[C_{\text{pn}}(\tau)J_{\text{ph}}(\tau)]d\tau,
\end{align}
 and the QD Lamb shift is $\Delta_{\rm Lamb} = \int_0^{\infty}\text{Im}[C_{\text{pn}}(\tau)J_{\text{ph}}(\tau)]d\tau$. Note that  the QD background decay, pure dephasing and incoherent pumping is also included in the ME. The resulting SE rate and the Lamb shift display an interplay between photon and phonon bath dynamics, where  $J_{\text{ph}}(\tau)$ and $C_{\text{pn}}(\tau)$ are the photon and the phonon bath correlation functions, respectively. The photon bath correlation function is defined as $J_{\text{ph}}(\tau) = \int_0^{\infty} d\omega J_{\text{ph}}(\omega) e^{i(\omega'_x-\omega)\tau}$, where the photon reservoir spectral function  is
\begin{align}
J_{\text{ph}}(\omega)= \frac{\v{d}\cdot \text{Im}[\v{G}(\r1_d,\r1_d;\omega)]\cdot \v{d}}{\pi\hbar\epsilon_0}.
\end{align}
The phonon correlation function 
\begin{align}
C_{\text{pn}}(\tau)=e^{[\phi(\tau)-\phi(0)]},
\end{align}
where the IBM phase function is defined through 
\begin{align}
\phi(t) = &\int_0^{\infty} d\omega\frac{J_{\rm pn}(\omega)}{\omega^2}
 \nonumber \\
&\left [\coth\left(\frac{\hbar\omega}{2k_BT}\right )\cos(\omega t)- i\sin(\omega t) \right],
\end{align}
which includes a sum over phonon emission and aborption processes.
Note that the SE rate $\tilde\gamma$ in principle includes contributions from photonic LDOS values at frequencies different from the ZPL  frequency of the QD (where $\omega = \omega_x'$). Such non-local frequency effects for the SE is caused by a breakdown of  Fermi's golden rule, which depends on the correlation times of both photon and phonon reservoirs~\cite{Kaushik}. We note that (\ref{eq5}) is broadly applicable irrespective of the specific structure of the photon reservoir. This approach  is restricted to weak-to-intermediate coupling between the QD and the photon reservoir, and thus it cannot treat the strong coupling regime which would manifest in effects such as vacuum Rabi spitting; however, this is generally the case for a photon ``reservoir.'' To treat
the strong coupling regime of quantum optics, e.g., for a high-Q cavity, one  includes a  cavity photon operator at the level of a system operator which requires a different approach based on the polaron cQED  ME that we describe below; while being able to describe the strong coupling regime, this approach is restricted (in the presence of phonon coupling) to model simple cavity structures and may also breakdown for low Q cavities~\cite{Kaushik}. 
\subsection{Cavity-QED polaron master equation theory}\label{sec2b} 

To derive the polaron cQED ME, we replace the photon  bath term in  the  Hamiltonian given by \eqref{eq1} with a single cavity mode, so that 
\begin{align}
\label{eq6}
    H &\!= \hbar \Delta_{cx}\adag\a  + \sum_q\hbar \omega_q b^{\dag}_qb_q
    +g(\smdag \a + \adag \sm)
     \nn\\
&\!+ (\sum_m \Omega_{m}a^\dagger a_m + {\rm H.c.} )+ \smdag\sm\sum_q\hbar\lambda_q(b^{\dag}_q+b_q),
\end{align}
where $\Delta_{cx} = \omega_c-\omega_x$ is the cavity-QD detuning, $g$ is the QD-cavity coupling for a single cavity mode described by lowering operator $a$, and $\Omega_m$ represents the coupling to the photon environment that causes
decay of the cavity mode. The time-dependence of $a$ is kept implicit in \eqref{eq6}. The QD-cavity coupling $g$ can be expressed in terms of the dipole moment $\v{d}=d\hat{\bf n}$ where $g = \eta({\bf n},{\bf r}_d)[\frac{d^2 \omega_c}{2\hbar \epsilon_0\epsilon V_{\rm eff}}]^{\frac{1}{2}}$, where $V_{\rm eff}$ is effective mode volume of a dielectric cavity with dielectric constant $\epsilon$, and
$\eta$ accounts for any deviation from the field antinode position and
misalignment in polarization coupling, and for optimal coupling
is simply 1.
  A dipole and a rotating wave approximation is used in describing the QD-cavity interaction. As before, we perform a polaron transform on $H$ and the polaron transformed form~\cite{Roy3} is now given by 
\begin{align}
\label{eq7}
    H' &= \hbar \Delta_{cx'}\adag\a  + \sum_q\hbar \omega_q b^{\dag}_qb_q \nn\\
&+g'(\smdag \a + \adag \sm)  + X_g\zeta_g+X_u\zeta_u,
\end{align}
where $g'=\avg{B}\!g$, $X_g = g[\smdag \a + \adag \sm]$, $X_u = ig[\smdag \a - \adag \sm]$ and the phonon fluctuation operators $\zeta_g = \frac{1}{2}(B_++B_--2\avg{B})$ and $\zeta_u=\frac{1}{2i}(B_+-B_-)$~\cite{Roy3} where $\avg{B} = \avg{B_+}=\avg{B_-}= e^{-\phi(0)/2}$. The polaron shift $\Delta_P$ is once again absorbed into the QD frequency $\omega'_x$ and $\Delta_{cx'} = \omega_c-\omega_x'$. We can then derive a time-convolutionless ME for the reduced density matrix $\rho$ of the QD-cavity system; following Ref.~\onlinecite{Roy3}, we use a second-order Born approximation with the polaron transformed interaction $H'_{\text I} = X_g\zeta_g+X_u\zeta_u$.
 The master equation in the interaction picture is then 
\begin{align}
\label{eq8}
   \frac{\partial \tilde\rho}{\partial t} = -\frac{1}{\hbar^2}\int_0^t dt' {\rm Tr_B}\{[\tilde H'_{\text I}(t),[\tilde H'_{\text I}(t'),\tilde \rho(t)\rho_{\rm B}]]\},
\end{align}
where $\tilde H'_{\text I}(t) = e^{i(H'_{\rm S}+H_{\rm B}')t/\hbar} H'_{\text I} e^{-i(H'_{\rm S}+H'_{\rm B})t/\hbar}$, with  $H'_{\rm S} = \hbar \Delta_{cx'}\adag\a   +g'[\smdag \a + \adag \sm]$
 and $H_{\rm B}'=\sum_q\hbar \omega_q b^{\dag}_qb_q$. The operator ${\rm Tr_B}$ denotes trace over the phonon bath ($\rho_{\rm B}$) and we assume the full density operator to be separable at all times $\rho\rho_{\rm B}$~\cite{Roy3}. Performing the trace and transforming back into the Schr\"{o}dinger picture, we obtain the final from of the time-local Markov polaron cQED ME~\cite{Roy3},
\begin{align}
\label{eq9}
&\frac{d \rho}{d t} = \frac{1}{i\hbar}[H'_{\rm S},\rho] +\frac{\gamma_0}{2}L(\sm)+\frac{\gamma_d}{2}L(\smdag\sm)
+\frac{P}{2}L(\smdag)\nn\\
&-\frac{1}{\hbar^2}\int_0^{\infty} d\tau \sum_{m = g,u}(G_m(\tau)[X_m,e^{-iH'_{\rm S}\tau/\hbar} X_m e^{iH'_{\rm S}\tau/\hbar}\rho(t)]\nn\\ &+\text{ H.c.}) + \frac{\kappa}{2}L(\a),
\end{align}
where $\kappa$ is the cavity decay rate, and
 $G_g= \avg{B}^2(\cosh(\phi(t))-1)$ and $G_u = \avg{B}^2 \sinh(\phi(t))$ are the polaron Green functions~\cite{Roy3}; as before, we have extended the upper limit of the integral in (\ref{eq8}) to $t\rightarrow \infty$ to obtain a Markov form. It should be noted that the incoherent processes are also included in the ME using the respective Lindblad terms. For the Born-Markov approximation to be valid, in the polaron frame, the system dynamics should be much slower compared to the rate of relaxation of the phonon bath. 
Although the bath relaxation time is only a few ps for typical QDs,
 this approximation may be restrictive in certain regimes, e.g., 
  when dealing with vacuum Rabi oscillations at large $g$ (Sec.~\ref{sec3b}) and off-resonant cavity feeding at large detunings $\Delta_{cx}$ (Sec.~\ref{sec3c}). In order to investigate this possible limitation, we extend a previously proposed phonon correlation expansion approach in Sec.~\ref{sec2c} by adding a cavity which is not limited by the Born-Markov approximation, though we only include phonon coupling to second-order. 
Although the correlation expansion includes phonon effects perturbatively,
the approach shows good agreement with the full IBM model for linearly excited bare QDs (i.e., with no photon bath) at low temperatures~\cite{Forstner}.
Photon coupling effects with the correlation expansion approach have been highlighted in Ref.~\onlinecite{Forstner3},
though the  exciton-photon correlation was  truncated using a Markov approximation, which is strictly valid only in weak coupling condition or for unstructured photon reservoirs.  
We highlight two useful limits of the 
cQED polaron ME: when phonons interactions are turned off, one fully
recover the Jaynes-Cummings model; and when the cavity is turned off,
one recovers the IBM~\cite{Imamoglu}, with the 
addition of important ZPL processes. 

Before concluding this section it should be noted that (\ref{eq9}) can be further simplified to a simpler effective master equation~\cite{Roy3}, if a weak excitation approximation (WEA) is made, which is exact for the linear spectrum. The cQED polaron ME now takes the analytical form, 
\begin{align}
\label{eq9a}
   &\frac{d \rho}{d t} = \frac{1}{i\hbar}[H^{\rm eff}_{\rm S},\rho] +\frac{\Gamma^{\smdag a}}{2}L(\smdag a)+\frac{\Gamma^{\adag \sm}}{2}L(\adag \sm)\nn\\
&+\frac{\gamma_0}{2}L(\sm)+\frac{\gamma_d}{2}L(\smdag\sm)+\frac{P}{2}L(\smdag)
  \nn\\
&+\gamma_{\rm cd}\adag \sm\rho\adag\sm +\gamma^*_{\rm cd}\smdag \a\rho\smdag\a \nn\\
&+  \left \{ M_1[(\adag\sm+\smdag\a),(2\smdag\sm\adag\a +\smdag\sm-\adag\a)\rho]  
+{\rm H.c.} \right \}\nn\\
&+\left \{ M_2[(\adag\sm-\smdag\a),(2\smdag\sm\adag\a +\smdag\sm-\adag\a)\rho]
+{\rm H.c.}\right\},
\end{align}
 where $H^{\rm eff}_{\rm S} = H'_{\rm S} +\hbar\Delta^{\adag\sm}\smdag\a\adag\sm+\hbar\Delta^{\smdag\a}\adag\sm\smdag\a$. If we denote the QD-cavity system Rabi frequency as $\Omega = \sqrt{\Delta_{cx'}^2+4g'^2}$
, then the phonon-mediated cavity/exciton scattering rates are given by
\begin{align}
&\Gamma^{a^\dag \sigma^-/\sigma^+ a}=  \nonumber \\
&2g'^2\text{Re}\left [\int_0^{\infty}d\tau \frac{2g'^2}{\Omega^2} (1-\cos(\Omega\tau))(e^{-\phi(\tau)} - 1) \right . \nonumber \\
&\left . +(\frac{2g'^2}{\Omega^2}\left ( 1-\cos(\Omega\tau))+\cos(\Omega\tau)\right)(e^{\phi(\tau)} - 1)\right] \nonumber \\
 &\pm 2g'^2\text{Im}\left [\int_0^{\infty}d\tau \frac{\Delta_{cx'}}{\Omega}\sin(\Omega\tau)(e^{\phi(\tau)} - 1)\right],
\end{align}
and the phonon-mediated Lamb shifts are given by 
\begin{align}
&\Delta^{a^\dag \sigma^-/\sigma^+ a}= \nonumber \\
&g'^2\text{Im}\left [\int_0^{\infty}d\tau \frac{2g'^2}{\Omega^2} (1-\cos(\Omega\tau))(e^{-\phi(\tau)} - 1) \right . \nonumber \\
&\left . +(\frac{2g'^2}{\Omega^2}(1-\cos(\Omega\tau))+\cos(\Omega\tau))(e^{\phi(\tau)} - 1)\right ] \nonumber \\
&\mp g'^2\text{Re}\left [\int_0^{\infty}d\tau \frac{\Delta_{cx'}}{\Omega}\sin(\Omega\tau)(e^{\phi(\tau)} - 1)\right ].
\end{align} 
We also have a cross-dephasing term~\cite{Ulhaq}
\begin{align}
& \gamma_{\rm cd}=
 2g'^2 \text{Re}\left [\int_0^{\infty}d\tau (\frac{2g'^2}{\Omega^2}(1-\cos(\Omega\tau)) \right .
\nonumber \\
& \left .+\cos(\Omega\tau)) (e^{-\phi(\tau)} - 1) 
+  \frac{2g'^2}{\Omega^2}(1-\cos(\Omega\tau)) (e^{\phi(\tau)} - 1)\right]
\nonumber \\
&-2ig'^2 \text{Re}\left [\int_0^{\infty}d\tau\frac{\Delta_{cx'}}{\Omega}\sin(\Omega\tau)(e^{-\phi(\tau)} - 1)\right],
\end{align} 
and the $M$ terms are defined through 
\begin{align}
M_1 & = -2g'^2 \int_0^{\infty}d\tau \frac{g'\Delta_{cx'}}{\Omega^2}(\cos(\Omega\tau)-1)
(\cosh{\phi}-1),\\
M_2 & = -2ig'^2 \int_0^{\infty}d\tau \frac{g'}{\Omega}\sin(\Omega\tau)\sinh{\phi}.
\end{align}

A physical understanding of the $M_{1,2}$ scattering terms can be obtained by deriving the Bloch equations for $\a$ and $\sm$, using the WEA. For example, 
\begin{align}
\label{eq20d}
\frac{d\avg{\a}}{dt} &= -g_c\sm-i(\Delta_{cx'}+\Delta^{\smdag\a})\a-\frac{\Gamma^{\rm eff}_c}{2}\a\nn\\
\frac{d\avg{\sm}}{dt} &= -g_x\a-i\Delta^{\adag\sm} \sm - \frac{\Gamma^{\rm eff}_x}{2}\sm,
\end{align}

\noindent where $\Gamma^{\rm eff}_c = \kappa+\Gamma^{\smdag a}$ and $\Gamma^{\rm eff}_x = \gamma_0+\gamma_d+P+\Gamma^{\adag\sm}$ are the effective dephasing rates and $g_c = ig'-M_1-M_2$ and $g_x = ig'+M_1-M_2$ are the complex couplings of the cavity and QD, respectively, in presence of phonons. Thus the processes denoted by the  scattering terms $M_{1,2}$, result in a complex coupling between QD and cavity, in the weak excitation approximation. At resonance ($\Delta_{cx'} = 0$) since $M_1$ = 0, so the complex QD-cavity coupling is given by $g_{c/x} = ig'-M_2$. 

Notably the above ME has solved the incoherent scattering terms exactly and one could use such an approach, e.g., to investigate the
strongly coupled spectra as a function of temperature.
The Rabi oscillations appearing in the integrals ensure that the phonon bath is correctly coupled to the dressed resonances of the system.

In the weak coupling limit, specifically when $\Delta_{cx'}\gg g'$, the Rabi frequency $\Omega \rightarrow \Delta_{cx'}$, and the incoherent cavity and exciton scattering rates are given by, ${\Gamma^{\adag \sm/\smdag \a}_0}=2g'^2 \text{Re}[\int_0^{\infty}d\tau e^{\mp i \Delta_{cx'}\tau}(e^{\phi(\tau)} - 1)]$ and phonon modified Lamb shifts are given by, ${\Delta^{\adag \sm/\smdag \a}_0}=g'^2 \text{Im}[\int_0^{\infty}d\tau e^{\mp i \Delta_{cx'}\tau}(e^{\phi(\tau)} - 1)]$. The phonon mediated scattering rates and Lamb shifts are the same as those derived in an earlier work~\cite{Roy3}, where an effective Lindblad form for the cavity-QED polaron ME was introduced.

A connection between the polaron reservoir ME approach (Sec.~\ref{sec2a}) and the polaron cQED theory can be made by expressing the SE rate (\ref{eq5}) in terms of these Lindblad decay rates~\cite{Kaushik}, 
\begin{align}
\tilde \gamma =\tilde \gamma_{\rm P} = \Gamma^{a^{\dagger} \sigma^-}_0+2g'^2\frac{(\frac{\kappa+\Gamma^{\smdag \a}_0-\Gamma^{\adag\sm}_0}{2})}{\Delta_{cx'}^2+(\frac{\kappa+\Gamma^{\smdag \a}_0
-\Gamma^{\adag\sm}_0}{2})^2},
\end{align}
which, however,   is only valid when  the spectral width of the cavity is much smaller than the width of the phonon bath function ($\approx$ 5 meV). This condition is satisfied by high Q cavities with $\kappa \le 0.1$ meV. Moreover, when $\kappa \gg \Gamma^{\smdag \a}_0-\Gamma^{\adag\sm}_0$, then\begin{align}
\label{eq9e}
 \tilde \gamma_{\rm P}  = \Gamma^{a^{\dagger} \sigma^-}_0+2g'^2\frac{(\frac{\kappa}{2})}
{\Delta_{cx'}^2+(\frac{\kappa}{2})^2},
\end{align}
which can be interpreted as a cavity-feeding term plus a phonon-modified (via $g\rightarrow g'$) cavity-induced SE  rate.

\subsection{Correlation expansion approach}\label{sec2c}
To derive a correlation expansion approach~\cite{Forstner, Forstner1}, we first derive the Heisenberg equations ($\dot{O} = -\frac{i}{\hbar}[O,H]$) for $\a$ and $\sm$ from (\ref{eq6}),  
\begin{align}
\label{eq10}
\dot{\a} &=-i\Delta_{cx} \a - i g\sm, \\
\dot{\sm} &= -i g\a - i\sum_q\lambda_q(\sm b^{\dag}_q +\sm b_q),
\end{align}
where we again make a WEA, i.e., $\sigma_z = -1$, to truncate the higher-order photon correlations. Again,  this is not restrictive in the current situation, since we excite the QD weakly to obtain the linear spectrum. The optical Bloch equations for $\avg{a}$ and $\avg{\sm}$ are obtained by performing ensemble average on the Heisenberg equations of motion, $\dot{\avg{O}} = -\frac{i}{\hbar}\avg{[O,H]}+{\rm Tr}[OL\rho]$,
giving 
\begin{align}
\label{eq11}
\dot{\avg{\a}} = -\left(i\Delta_{cx} +\frac{\kappa}{2}\right )\avg{\a} - i g\avg{\sm},\nn\\
\dot{\avg{\sm}} = -i g\avg{\a} -\left(\frac{P+\gamma_0+\gamma_d}{2}\right)\avg{\sm}\nn\\
- i\sum_q\lambda_q(\avg{\sm b^{\dag}_q} +\avg{\sm b_q}),
\end{align}
where the Lindblad operators for background radiative decay, pure dephasing and incoherent pumping have been included. The time-evolution of  $\avg{\sm}$ depends on phonon-assisted correlations $\avg{\sm b_q}$, whose equations are given by
\begin{align}
\label{eq12}
\dot{\avg{ \sm b_q}} &= -ig \avg{\a b_q}  - \left(i \omega_q+\frac{P+\gamma_0+\gamma_d}
{2}\right) \avg{ \sm b_q}  \nn\\
&- i\sum_m\lambda_m\avg{\sm b_q b_m} 
- i\sum_m\lambda_m \avg{\sm b_q b^{\dag}_m },\nn\\
\dot{\avg{ \sm b_q}}_c &= -ig \avg{\a b_q}_c  - 
\left(i \omega_q + \frac{P+\gamma_0+\gamma_d}{2}\right) \avg{ \sm b_q}_{\rm c}\nn\\  
&- i\lambda_q (n_q+1)\sm\nn\\
&- i\sum_m\lambda_m\avg{\sm b_q b_m}_{\rm c}
- i\sum_m\lambda_m \avg{\sm b_q b^{\dag}_m }_{\rm c},
\end{align}
 where $n_q = (e^{\frac{\hbar \omega_q}{K_B T}} -1)^{-1}$ is the average phonon occupation number for mode $q$ given by the thermal equilibrium Bose-Einstein distribution at temperature $T$. The simplification in the second line of \eqref{eq12} is performed using the correlation expansion technique~\cite{Forstner1}. The operator average $\avg{\sm b_q b^{\dag}_m}$ can be expanded as $\avg{\sm b_q b^{\dag}_m} = \avg{\sm} \avg{b_q} \avg{b^{\dag}_m} + \avg{\sm b_q} \avg{b^{\dag}_m} + \avg{\sm b^{\dag}_m} \avg{b_q} + \avg{\sm} \avg{b_q b^{\dag}_m} + \avg{\sm b_q b^{\dag}_m}_{\rm c}$, where the three operator average is separated using cluster expansion techniques into singlets ($\avg{O}$), doublet operator averages ($\avg{O O'}$)  and a correlated part $\avg{O O' O^{''}}_{\rm c}$. The pure phonon correlations such as $\avg{b_q b^{\dag}_m}_{q\neq m}$ and derivatives as $\dot{\avg{b_q}}$, $\dot{\avg{b^{\dag}_qb_q}}$ are neglected, as they represent phonon coherences out of equilibrium and the approximation is similar to a bath approximation~\cite{Forstner1}. Thus $\avg{b^{\dag}_qb_q} = n_q$  and $\avg{b^{\dag}_q} =\avg{b_q} = 0$ and we have $\avg{\sm b_q b^{\dag}_m} =   \avg{\sm} \avg{b_q b^{\dag}_m} + \avg{\sm b_q b^{\dag}_m}_{\rm c}$.  We next observe that $\avg{\sm b_q}$ depends on 
higher-order phonon-assisted correlations such as $\avg{\sigma^+ b_q b_m}_{\rm c}$ and $\avg{\sm b_q b^{\dag}_m}_{\rm c}$ and this leads to a hierarchy of coupled equations. In the current work, the hierarchy is truncated by dropping all correlations involving three or more phonon  operators, such as $\avg{\sm b_q b_m b_n}_{\rm c}$. Thus, using this truncation approximation the equations for second-order phonon assisted correlations are given by 
\begin{align}
\label{eq13}
&\dot{\avg{\sm b_q b_m}}_{\rm c} = - ig \avg{\a b_q b_m}_c\nn\\ &
-\left(i(\omega_q+\omega_m)+\frac{P+\gamma_0+\gamma_d}{2}\right)\avg{\sm b_q b_m}_{\rm c} \nn\\
&-i\avg{\sm b_q}_c\lambda_m(n_m+1) - i\avg{\sm b_m}_{\rm c}\lambda_q(n_q+1).
\end{align}
 The  equations of motion for the phonon-assisted cavity correlation are as follows:
\begin{align}
\label{eq14}
\dot{\avg{\a b_q}_{\rm c}} &= -\left (i(\Delta_{cx} + \omega_q) +\frac{\kappa}{2}\right) \avg{\a b_q}_{\rm c} -i g \avg{\sm b_q}_{\rm c}, \\
\dot{\avg{\a b_q b_m}_{\rm c}} &= -\left(i(\Delta_{cx} + \omega_q+\omega_m)
+\frac{\kappa}{2}\right) \avg{\a b_q b_m}_{\rm c} \nn\\
&-i g \avg{\sm b_q b_m}_{\rm c}. 
\end{align}
The truncation approximation helps us to derive a closed set of equations in the one-time operator averages ($\avg{\sm}$, $\avg{a}$), first order ($\avg{\sm b_q}_{\rm c}$, $\avg{\a b_q}_{\rm c}$) and second-order ($\avg{\sm b_q b_m}_{\rm c}$, $\avg{\a b_q b_m}_{\rm c}$) phonon correlations, which can be solved using direct numerical integration. The emission spectrum (see Sec.~\ref{sec2d}) requires the two-time operator averages or correlation function as $\avg{\smdag(t)\sm(t+\tau)}$ and $\avg{\adag(t)\a(t+\tau)}$. These two-time correlations can be calculated using the one-time coupled equations and the quantum regression theorem~\cite{Carmichael}. The quantum regression theorem states that if single time operator averages for a set of operators $O_n(t)$ satisfy a closed set of linear coupled equations as $\frac{d\avg{O_n}}{dt} = \mathcal{L} (\avg{O_m})$ (where $\mathcal{L}$ denotes a linear combination of $O_n$), then a two-time average such as $P(t)O_n(t+\tau)$ involving $O_n$ will also evolve obeying the same set of equations, $\frac{d\avg{P(t)O_n(t+\tau)}}{d\tau} = \mathcal{L} (\avg{P(t)O_m(t+\tau)} $). For example, the equation for the correlation function $\avg{\smdag(t)\sm(t+\tau)}$ is obtained using the regression theorem in \eqref{eq11} as
\begin{align}
\label{eq15}
\frac{d \avg{\smdag(t)\sm(t+\tau)}} {d\tau} 
&= -i g\avg{\smdag(t)\a(t+\tau)} \nn\\&-\frac{P+\gamma_0+\gamma_d}{2}\avg{\smdag(t)\sm(t+\tau)}\nn\\
&\!\!\!\!\!- i\sum_q\lambda_q  \left (\avg{\smdag(t) \sm(t+\tau) b^{\dag}_q(t+\tau)}_{\rm c}  \right . \nn\\&
\left . 
\phantom{\sum\!\!\!\!\!\!\!\!\!\!\!} 
\!\!\!\!\!+\avg{\smdag(t) \sm(t+\tau) b_q(t+\tau)}_{\rm c} \right ).
\end{align}

\noindent Similar equations can be derived for other two-time operator averages to form a closed set, which can be numerically integrated starting from proper initial conditions given by quantities such as $\avg{\smdag (t_s)\sm(t_s)}$, $\avg{\smdag(t_s)\sm(t_s) b_q(t_s)}_{\rm c}$, $\avg{\smdag(t_s)\a(t_s)}$, where $t_s$ denotes the time when the system reaches steady state. The above steady-state expectation values can be evaluated using  Bloch equations for these expectation values. The expectation values are also determined under the two-phonon correlation expansion approximation. As shown below (Sec.~\ref{sec2d}), these correlation functions will be used for calculating the WEA emission spectra and are accurate at low temperatures~\cite{Forstner}.

Finally, it should be noted that a polaron shift $\Delta_{\rm P}$ of the QD frequency ($\omega_x$) automatically appears in the spectrum calculated from the correlation expansion approach and, for convenience, can again be absorbed into the polaron shifted QD frequency ($\omega'_x$).




\subsection{Emission spectrum}\label{sec2d}

\subsubsection{Exact expression for the photon Green function spectrum for a generalized photon reservoir}\label{sec2d1}

The incoherent emission spectrum of a QD coupled to a structured reservoir at a point detector at position $\r1_{\rm D}$ is exactly given by~\cite{Ge}
\begin{align}
\label{eq15a}
S^{\rm G} (\r1_{\rm D},\omega) 
& = \avg{(\v{E}^+(\r1_{\rm D},\omega))^{\dagger}\,\v{E^+(\r1_{\rm D},\omega)}},
\end{align}
where $\v{E}^+(\r1_{\rm D},\omega) = \int^{\infty}_0 dt e^{i\omega t}\v{E}^+(\r1_{\rm D},t)$ is the Laplace transform of the  positive frequency component of the total electric field operator  $\v{E}(\r1_{\rm D},t),$ defined as 
\begin{align}
\label{eq15b}
&\v{E}(\r1_{\rm D},t) = \v{E}^+(\r1_{\rm D},t) + \v{E}^-(\r1_{\rm D},t)\nn\\
&=\int^{\infty}_0 d\omega' [\v{E}^+(\r1_{\rm D},\omega',t) + \v{E}^-(\r1_{\rm D},\omega',t)]\nn\\
&=i\int^{\infty}_0 d\omega' \int d\r1' \v{G}(\r1_{\rm D},\r1';\omega')\sqrt{\frac{\hbar}{\pi\epsilon_0}\epsilon_I(\r1',\omega')}\f(\r1',\omega',t) \nn\\&+\rm{H.c.}
\end{align}
In the frequency domain,  
\begin{align}
\label{eq15c}
\v{E}^+(\r1_{\rm D},\omega) &= i\int^{\infty}_0 d\omega' \int d\r1' \v{G}(\r1_{\rm D},\r1';\omega')\nn\\
&\times \sqrt{\frac{\hbar}{\pi\epsilon_0}\epsilon_I(\r1',\omega')}\f(\r1',\omega',\omega).
\end{align}
Starting from the original Hamiltonian $H$ in (\ref{eq1}), the electric field operator can be expressed in terms of the QD polarization using Laplace transform techniques, yielding~\cite{Ge}
\begin{align}
\label{eq15d}
\v{E}(\r1_{\rm D},\omega) = \v{E}^0(\r1_{\rm D},\omega) + \v{G}(\r1_{\rm D},\r1_d,\omega)\cdot \v{d}\frac{\sm(\omega)}{\epsilon_0},
\end{align}
 where $\v{E}^0$ denotes the free-field solution in the absence of a QD scatterer. The medium Green function $\v{G}(\r1_{\rm D},\r1_d,\omega)$  includes all propagation effects~\cite{Ge}, including radiative losses due to structured reservoirs~\cite{Kaushik}; e.g., in the case of a planar geometry, this would also include out of plane radiative losses to photonic modes above the light line~\cite{Hughes2}. In the following treatment, we assume that the Green function only accounts for radiative losses due to the structured photonic reservoir. Any small additional out of plane radiative losses is accounted for by the phenomenological Lindblad term $\gamma_0$. Assuming the initial vacuum state of the photonic reservoir, the incoherent spectrum from Green function theory can be derived using (\ref{eq15a}) and (\ref{eq15d}), giving 
\begin{align}
\label{eq16}
S^{\rm G}(\r1_{\rm D},\omega) &= \abs{\frac{\v{G}(\r1_{\rm D},\r1_d;\omega)\cdot \v{d}}{\epsilon_0}}^2\avg{\smdag(\omega)\sm(\omega)} \nn\\
&= \alpha_{\rm prop}({\bf r}_{\rm D},{\bf r}_{d};\omega)\, S_0(\omega),
\end{align}
\noindent where $S_0=\avg{\smdag(\omega)\sm(\omega)}$ is the polarization spectrum and $\alpha_{\rm prop} = \frac{1}{\epsilon_0}\abs{\v{d}\cdot\v{G}(\R_{\rm D},\r1_d;\omega)}^2$ accounts for propagation and filtering from the QD
(at ${\bf r}_d$) to the detector (at ${\bf r}_{\rm D}$). The polarization spectrum can be written as
\begin{align}
\label{eq16a}
S_0(\omega) &=\avg{\smdag(\omega)\sm(\omega)}\nn\\
&=\avg{ \left (\int^{\infty}_0 dt_1 e^{i\omega t_1} \sm(t_1)\right)^{\dagger} \int^{\infty}_0 dt_2 e^{i\omega t_2} \sm(t_2)} \nn\\
&=\int^{\infty}_0 dt_1 \int^{\infty}_0 dt_2 e^{i\omega (t_2-t_1)} \avg{\smdag(t_1)\sm(t_2)}.
\end{align}
\noindent Using a rotating frame at the exciton frequency ($\omega'_x$)  denoting $\tau = t_2 -t_1$ and taking the limit of $t_1 = t\rightarrow\infty$, we derive the steady-state polarization spectrum, 
\begin{align}
\label{eq17}
S_0(\omega) &= \lim_{t\rightarrow\infty}\text{Re}\left[\int_0^{\infty}d\tau\avg{\smdag(t+\tau)\sm(t)}
e^{i(\omega'_x-\omega)\tau}\right],
\end{align}
 which can be calculated using the techniques developed in the previous sections. When calculated using the polaronic approach (Secs.~\ref{sec2a}, \ref{sec2b}), a transformation is required to obtain $S_0$ in the lab frame. If $\sm_{\rm P}$ denotes the QD lowering operator in the polaron frame, then a transformation of the correlation function ($\avg{\smdag_{\rm P}(t+\tau)\sm_{\rm P}(t)}$) from the polaron to the lab frame,
\begin{align}
\label{eq17a}
 \avg{\smdag_{\rm P}(t+\tau)\sm_{\rm P}(t)} \rightarrow \avg{\smdag(t+\tau)B_+(t+\tau)B_-(t)\sm(t)}, 
\end{align}
 which produces a phase relaxation term $e^{\phi(\tau)}$~\cite{Roy3} that accounts for the phonon-induced pure dephasing of the QD polarization.
The overall decay is thus clearly non-Markovian and this is a strong advantage of the polaronic approaches, i.e., even though they use a Born-Markov approximation
for the equations of motion, non-Markovian coupling to the phonon reservoir are captured through the polaron transform (indeed, the approach fully includes the IBM solution).
 The final polarization spectrum calculated using the polaron approach is then $S_0(\omega) = \lim_{t\rightarrow\infty}\text{Re}[\int_0^{\infty}d\tau\avg{\smdag_{\rm P}(t+\tau)\sm_{\rm P}(t)}e^{\phi(\tau)}e^{i(\omega'_x-\omega)\tau}]$. The expression for the emission spectrum (\ref{eq16}) is exact and only limited by the approximations made in calculation of $S_0$ from the above theories.
Thus  no phenomenological input-output formalism is required here as the Green function is already a solution for the scattering problem and the propagation of light in the medium. In order to distinguish $S^{\rm G}$ computed from the several different approaches for obtaining the system dynamics, we identify the following spectra: $S^{\rm G-res}_{\rm cav/crow}$, $S^{\rm G-cQED}_{\rm cav}$ and $S^{\rm G-ce}_{\rm cav}$, when $S_0$ is calculated using reservoir polaron ME,  cQED polaron ME and the correlation expansion approach, respectively;
the  subscripts refer to a cavity or CROW  medium as examples of the photonic bath, but note only the former approach can model arbitrary  photon bath spectral functions (e.g., a non-Lorentzian lineshape).

The  propagator function $\alpha_{\rm prop}(\omega)$ filters the polarization spectra $S_0$  as the light propagated from the QD to the detector, and depends on the specific structure of the reservoir. For a single mode cavity with Lorentzian broadening, it is given by~\cite{Hughes2}
\begin{align}
\label{eq18}
\alpha_{\rm prop}^{\rm cav}(\omega) &=\abs{ \frac{ \v{f}_{ c}(\r1_{\rm D}) \otimes \v{f}^*_{ c}(\r1_{ d})\cdot \v{d} } {\epsilon_0} }^2 \abs{\frac{\omega_{\rm c}^2}{\omega^2_{c}-\omega^2 - i\kappa\omega}}^2,
\end{align}
where $\v{f}_{ c}(\r1)$ is the normalized transverse  mode of the cavity.  In the case of a high-Q cavity,
and assuming $\kappa \ll \omega_c$,
 (\ref{eq18}) reduces to a simple Lorentzian lineshape and the resulting Green function emission spectrum is given by 
\begin{align}
\label{eq18a}
S^{\rm G}_{\rm cav}(\r1_{\rm D},\omega) &= 
\alpha^{\rm cav}_{0}\frac{\frac{\kappa}{2}}{(\omega-\omega_{ c})^2 + (\frac{\kappa}{2})^2} \avg{\smdag(\omega)\sm(\omega)},
\end{align}
\noindent where  $\alpha^{\rm cav}_{0} = \abs{ \frac{ \v{f}_{ c}(\r1_{\rm D}) \otimes \v{f}_{c}^*(\r1_{\rm d})\cdot\v{d} } {\epsilon_0} }^2\frac{\omega_{ c}^2}{2\kappa}$ denotes the frequency-independent prefactor.

For a slow-light coupled-cavity waveguide (CROW)~\cite{Yariv},  the frequency-dependent propagation factor is 
\begin{align}
\label{eq19}
\alpha_{\rm prop}^{\rm crow}(\omega) &= \abs{ \frac{ \v{f}_{\rm crow}(\r1_{\rm D}) \otimes \v{f}^*_{\rm crow}(\r1_{\rm d})\cdot \v{d} } {\epsilon_0} }^2  \frac{\omega^2}{4}
\times
\nn\\ &\abs{\frac{1}{ \sqrt{ (\omega-\tilde\omega_{\rm l}^*) (\omega-\tilde\omega_{\rm u}) }}}^2,
\end{align}
where $\tilde\omega_{\rm u,l} = \omega_{\rm u,l} +\kappa_{\rm u,l}$, and $\omega_{\rm u,l}$ marks the upper and  lower mode-edge frequencies of the waveguide~\cite{Kaushik}, $\kappa_{\rm u,l}$ denotes damping and $\v{f}^{\rm }_{\rm crow}(\r1)$ is the normalized mode of the  cavity containing the QD in the CROW\cite{Fussell1, Fussell2}. The Green function emission spectrum from the waveguide is then
\begin{align}
\label{eq19a}
S^{\rm G-res}_{\rm crow}(\r1_{\rm D},\omega) &= \alpha^{\rm crow}_0 \abs{\frac{1}{ \omega\sqrt{ (\omega-\tilde\omega_{\rm l}^*) (\omega-\tilde\omega_{\rm u}) }}}^2 \times \nn\\
&\avg{\smdag(\omega)\sm(\omega)}.
\end{align}
\noindent where $\alpha^{\rm crow}_0 = \abs{ \frac{ \v{f}_{\rm crow}(\r1_{\rm D}) \otimes \v{f}^*_{\rm crow}(\r1_{\rm d})\cdot \v{d} } 
{\epsilon_0} }^2$. For a fixed detector location, the frequency-independent pre-factors $\alpha^{\rm cav/crow}_{\rm 0}$  containing the spatial mode functions do not influence the normalized emission spectrum.

\subsubsection{Coupled mode approach with correlation expansion and a simple coupled cavity}\label{sec2d2}

In the case where the structured reservoir consists of a Lorentzian cavity, the dynamics of the photon reservoir can be approximately described by a single mode lowering operator $a$ (strictly valid in high Q case) in the system Hamiltonian $H$ (\ref{eq6}). The damping of the cavity mode to the environment is described using the phenomenological decay $\kappa$.  In this case, a calculation of the reservoir/cavity emitted spectrum can be obtained using 
\begin{align}
\label{eq20}
S^{\rm CM}_{\rm cav}(\omega) &=
F(\r1_{\rm d},{\bf r}_{\rm D})\frac{\kappa}{\pi}\times \nonumber \\ &\lim_{t\rightarrow\infty}
\text{Re}\left[\int_0^{\infty}d\tau\avg{a^{\dagger}(t+\tau)a(t)}e^{i(\omega_x'-\omega)\tau}\right]\nn\\
&=F(\r1_{\rm d},{\bf r}_{\rm D})\frac{\kappa}{\pi}\avg{(a(\omega))^{\dagger}a(\omega)},
\end{align}
where $a(\omega)$ is the Laplace transform of $a(t)$,
and $F(\r1_{\rm d},{\bf r}_{\rm D})$ is a frequency-independent geometrical factor to account for 
the propagation from the QD to the detector position.
The superscript $\rm CM$ denotes coupled mode formalism which treats the QD and cavity as coupled modes and the photonic reservoir in this case is assumed to be described a single cavity mode $a$. The structure (width) of the cavity determines its damping rate and any such reservoir effects arising due to finite cavity lifetime is incorporated using phenomenological decay terms $\kappa$ for the cavity operator. An analytical expression for $a(\omega)$ can be derived using the Heisenberg equations for $a$. Starting from the coupled mode system Hamiltonian $H$ (\ref{eq6}), the Heisenberg equation for the cavity mode operator is 
\begin{align}\label{eq20a}
\frac{da}{dt} &= -\left (i\Delta_{cx}+ \frac{\kappa}{2}\right )a -ig\sm,
\end{align}
where the phenomenological cavity decay is included. Laplace transforming the $a$ equation gives
$a(\omega)= a^0 +\frac{ig}{(\omega-\omega_c)+i\frac{\kappa}{2}}\sm(\omega)$, where $a^0$ is the free field. Assuming an initial vacuum state for the cavity and substituting $a(\omega)$ in coupled mode spectra $S^{\rm CM}_{\rm cav}(\omega)$, we derive
\begin{align}
\label{eq20b}
S^{\rm CM}_{\rm cav}(\omega) &=F({\bf r}_D)\frac{\frac{g^2}{\pi}}{(\omega-\omega_c)^2 + (\frac{\kappa}{2})^2} \avg{\smdag(\omega)\sm(\omega)}.
\end{align}
This functional form of the coupled-mode spectrum $S^{\rm CM}_{\rm cav}$ is similar to the Green function spectrum (\ref{eq18a}). Hence for a fixed position of detector, the normalized spectrum derived from these two theories should match closely in the case of a Lorentzian cavity reservoir. This, however, happens only if all the approximations used in the calculations of $\avg{\smdag(\omega)\sm(\omega)}$ and $\avg{\adag(\omega)\a(\omega)}$ are appropriate and consistent. For example, the polaron cQED approach is derived using the Born-Markov approximation with polaron transformed interaction terms. When applied beyond the range of the Born-Markov approximation, the coupled-mode and Green function spectra may lead to different results (shown later). Hence,  we will use this models to investigate the validity of the different theories in Secs.~\ref{sec3b} and \ref{sec3c}. The two approaches where the coupled-mode spectrum can be computed are polaron cQED ME and correlation expansion approach. To distinguish $S^{\rm CM}_{\rm cav}$ calculated using correlation expansion and polaron cQED ME, we label these as $S^{\rm CM-ce}_{\rm cav}$ and $S^{\rm CM-cQED}_{\rm cav}$, respectively. We discuss the latter coupled mode spectra in details below.

\subsubsection{Coupled mode approach with the cavity-QED polaron master equation}\label{sec2d3}

The coupled mode spectrum (i.e., not using the exact Green function expression) in the polaron frame using the polaron cQED ME approach is given by 
\begin{align}
\label{eq20c}
&S^{\rm CM-cQED}_{\rm cav}(\omega) =F({\bf r}_D)\frac{\kappa}{\pi}\avg{(a_{\rm P}(\omega))^{\dagger}a_{\rm P}(\omega)}=  \nn\\
&\,\,\,\,\,\frac{F({\bf r}_D) \kappa}{\pi}\lim_{t\rightarrow\infty} 
\text{Re}\left[\int_0^{\infty}d\tau\avg{a^{\dagger}_{\rm P}(t+\tau)a_{\rm P}(t)}e^{i(\omega_x'-\omega)\tau}\right],
\end{align}
\noindent where $a_{\rm P}$ is the cavity lowering operator in the polaron frame. However the polaron transform does not influence the cavity operator $a = a_{\rm P}= e^P a e^{-P}$, since $P = \smdag\sm \sum_q\frac{\lambda_q}{\omega_q} (b^{\dag}_q-b_q)$~\cite{Roy3}. Thus the coupled mode spectrum remains unchanged when a transformation is made from the polaron to the lab frame 
(i.e.,$\avg{(a_{\rm P}(\omega))^{\dagger}a_{\rm P}(\omega)} = \avg{(a(\omega))^{\dagger}a(\omega)}$).
 
A simple analytic expression for the coupled mode spectra $S^{\rm CM-cQED}_{\rm cav}(\omega)$ can be derived in the WEA using the simplified cQED polaron ME ~\eqref{eq9a}. We start from Bloch equations (\ref{eq20d}) in $\avg{a}$ and $\avg{\sm}$. The equations for the two-time correlations $\avg{\smdag(t)\sm(t+\tau)}$ and $\avg{\adag(t)\a(t+\tau)}$ are the same as (\ref{eq20d}) from the quantum regression theorem~\cite{Carmichael}. These equations are subsequently Laplace transformed~\cite{Yao1} and solved to give the cavity-emitted  spectrum

\begin{align}
\label{eq20e}
S^{\rm CM-cQED}_{\rm cav} = \frac{i\avg{\adag\a}_{ss}D(\omega)+g_c\avg{\adag\sm}_{\rm ss} }{D(\omega)C(\omega)+g_cg_x},
\end{align}
where $D(\omega)=(\omega-\Delta^{\adag\sm})
+i\frac{\Gamma^{\rm eff}_x}{2}$ , $C(\omega)=(\omega-\Delta_{cx'}
-\Delta^{\smdag \a})+i\frac{\Gamma^{\rm eff}_c}{2}$, and the steady-state expectation values $\avg{\adag a}_{ss}$ and $\avg{\adag \sm}_{\rm ss}$ are calculated using Bloch equations for $\avg{\adag a}_{\rm ss}$, $\avg{\smdag\sm}_{\rm ss}$, $\avg{\adag \sm}_{\rm ss}$ and $\avg{\smdag \a}_{\rm ss}$. One obtains 
\begin{align}
\label{eq20f}
\avg{\adag\a}_{\rm ss}\!=\!\frac{P( (\Delta'^2-\abs{\gamma_{\rm cd}}^2 + \frac{\Gamma^2_{\rm Total}}{4}) \Gamma^{\adag\sm} \!+\!2{\rm Re}[g_1N_2]  )}{\bf D},
\end{align}
\begin{align}
\label{eq20f_g}
\avg{\adag\sm}_{\rm ss} = \frac{(N_1(P+\gamma_0)-\kappa N_2)\avg{\adag\a}_{\rm ss} + PN_2}{(P+\gamma_0) (\Delta'^2-\abs{\gamma_{\rm cd}}^2 + (\frac{\Gamma_{\rm Total}}{2})^2 )},
\end{align}
 where ${\bf D} =(\Delta'^2-\abs{\gamma_{\rm cd}}^2 + \frac{\Gamma^2_{\rm Total}}{4})(\Gamma^{\rm eff}_c(P+\gamma_0) + \kappa \Gamma^{\adag\sm})-2{\rm Re}[g_1(N_1(P+\gamma_0)-\kappa N_2)]$, and  $\Gamma_{\rm Total} = \Gamma^{\rm eff}_x + \Gamma^{\rm eff}_c$ is the total dephasing rate and $\Delta' =\Delta_{cx'}+\Delta^{\smdag\a}- \Delta^{\adag\sm}$ is the net QD-cavity detuning; also, $g_1 = 2M^r_1-i(g'-2M^i_2)$ where $M_1 = M^r_1+iM^i_1$ and $M_2 = M^r_2+iM^i_2$ ($r$ and $i$ superscript denotes real and imaginary parts, respectively); while the factors, $N_1 = \gamma^*_{\rm cd}g^*_3 + g_3(\frac{\Gamma_{\rm Total}}{2}+i\Delta')$ and $N_2 = \gamma^*_{\rm cd}g^*_4 + g_4(\frac{\Gamma_{\rm Total}}{2}+i\Delta')$, where $g_3 = 2M^r_2-i(g'+2M^i_1)$ and $g_4 = 2M^r_2+i(g'-2M^i_1)$.

%
%
%
%
%

%

\subsubsection{Cavity-emitted spectrum from
a semiclassical linear susceptibility theory}\label{sec2d4}

 In this section we   discuss the cavity-emitted spectrum derived using the linear susceptibility theory~\cite{HughesSC,savonaPRB}. The linear susceptibility approach is semiclassical and has been previously used for analyzing emission spectrum of a coupled QD-cavity system,  specifically for investigating the role of phonons on the  asymmetric vacuum Rabi-doublet\cite{HughesSC} and for connecting to experiments on off-resonant cavity feeding~\cite{savonaPRB,HughesSC}; but its predictions and assumptions have never been compared against the  quantum optical approaches starting with the full Hamiltonian (Secs.~\ref{sec2a}, \ref{sec2b} and \ref{sec2c}). In order to carry out such a comparison, here we briefly review the  linear susceptibility approach. The theory assumes a time-dependent QD polarizability as $p_x(t) = p_x(0) e^{-i\omega'_xt-\frac{\Gamma_x}{2}t +\phi(t)}$, where $\Gamma_x = \gamma_0+\gamma_d$ is the total dephasing rate of the bare QD and $p_x(0)$ is a constant prefactor. In order to incorporate phonon effects to all orders, the time-dependent IBM phase $\phi$ is added to the Lorentzian decay of bare QD in the expression of $p_x(t)$. Next $p_x(t)$ is  Fourier transformed to generate the frequency-dependent linear polarizability/susceptibility, $\chi(\omega) = 2\omega'_x/((\omega'_x)^2-\omega^2-i\omega\Gamma_x-\omega\Sigma_{\rm ph}(\omega) )$, where $\Sigma_{\rm ph}(\omega)$ is the phonon self-energy. In the absence of a photonic reservoir, $\chi$ describes the linear response of a QD in presence of phonons to a weak pump. The imaginery part of $\chi$ gives a measure of the linear absorption spectrum, which has the IBM spectral line-shape~\cite{HughesSC,savonaPRB} (see Fig.~\ref{fig3}, dashed line). Coupling to a structured photonic reservoir is included in this theory by assuming photon and phonon correlations to be independent. Thus the photonic reservoir only produces a SE enhancement (Purcell effect) and a photonic Lamb shift. For a single Lorentzian cavity (low or high Q), the QD susceptibility is
\begin{align}
\label{eq20g}
\chi(\omega) = \frac{2\omega'_x}{( (\omega'_x)^2-\omega^2-i\omega\Gamma_x-\omega\Sigma(\omega) ) -\frac{4g^2\omega'_x\omega_c}{( \omega_c^2-\omega^2-i\omega\kappa )}}.
\end{align}
Now assuming that the exciton is initially excited, 
 the cavity-emitted spectrum is derived to be~\cite{HughesSC,savonaPRB}
\begin{align}
\label{eq20h}
S^{\rm G-sus}_{\rm cav} &=\kappa F(\r1_{\rm d},{\bf r}_{\rm D}) \times \nn\\  &  
\abs{\frac{\frac{2g\omega_c(\omega+\omega'_x)}{( \omega_c^2-\omega^2-i\omega\kappa )}}{( (\omega'_x)^2-\omega^2-i\omega\Gamma_x-\omega\Sigma(\omega) ) -\frac{4g^2\omega'_x\omega_c}{( \omega_c^2-\omega^2-i\omega\kappa )}}}^2.
\end{align}
 where $F(\r1_{\rm d},{\bf r}_{\rm D})$ denotes the frequency-independent geometrical factor that depends on the QD and detector position~\cite{Yao1}. The spectrum can be approximated as a projection of QD susceptibility $\chi(\omega)$ (\ref{eq20g}) by the Lorentzian cavity function 
(\ref{eq18})~\cite{savonaPRB,HughesSC} and is hence identified with the superscript ${\rm G}$ of the Green function projected spectrum.  In the absence of phonon coupling (but accounting for the ZPL decay), the spectrum has the form of two coupled Lorentzians. With phonon coupling, the QD component becomes non-Lorentzian due to presence of a complex self-energy. 

 Though applied in the case of a simple cavity here, the semiclassical linear susceptibility theory, in principle, can be generalized to arbitrary photonic reservoirs~\cite{HughesSC}. The theory also does not make any Markov approximations. However the self-energy $\Sigma(\omega)$ derived using this approach depends on the QD dephasing rate $\Gamma_x$, which means that the model not self-consistent. The theory also cannot distinguish between physically distinct  pure dephasing and radiative damping processes, though (using a ME approach) these are found to affect the linear  spectra in an identical manner. The formalism also lacks a clear incoherent pumping scheme and assumes an initially excited QD (i.e., $\avg{\smdag\sm}(t=0) =0$). This assumption however makes no difference in the calculated spectra as long as the WEA is valid (see Appendix~\ref{Appen1} for details). Finally, the susceptibility theory  cannot model any non-linear optical effects, which can be easily included in polaronic ME approaches (though the phonon-mediated scattering rates are no longer analytically solvable in general). 

 At this point we have now introduced  the various formalisms that can be used for calculating the emission spectra of excited QDs coupled to structured photonic reservoirs in the presence of acoustic phonons. Before discussing the results, we next introduce the specific phonon reservoir parameters which will be used in the subsequent calculations and we also discuss the computational pros and cons of the different numerical schemes.

\subsection{Phonon model and QD parameters}\label{sec2e}
So far in the theory we have not made any specific assumption about the structure of the phonon bath.  For small epitaxial  semiconductor QDs, the dominant phonon interaction arises due to deformation potential coupling to LA phonon modes~\cite{Roy2}.
The expression for phonon coupling $\lambda_q$ for deformation potential coupling to LA phonons is $\lambda_q = \sqrt{\frac{\hbar \omega_q}{2\rho_{\rm d} c^2_{\rm l} V}} \frac{D}{\hbar}e^{-\frac{\omega^2}{4\omega^2_{\rm p}}}$~\cite{Forstner1} where $\rho_{\rm d}$ is the material density, $V$ is the material volume, $c_{\rm s}$ is the sound velocity, $D=D_{\rm val}-D_{\rm con}$ is the difference in deformation potential between the valence  and conduction  bands and $\omega_{\rm p}$ is the phonon cut-off frequency determined by the confinement length of the electron. The sum over the discrete phonon modes $\lambda_q$ can be converted into an integral using the density of states of phonons $D(\omega) = \frac{V}{(2\pi)^3}\frac{4\pi\omega^2}{c^3_{\rm s}}$ as follows,
\begin{align}
\label{eq21}
\sum_q \lambda_q &\rightarrow \int_0^{\infty} d\omega D(\omega) \lambda (\omega)\nn\\
& = \int_0^{\infty} d\omega \sqrt{V} \omega  \sqrt{J_{\rm pn}(\omega)} \frac{1}{\sqrt{2\pi^2c^3_{\rm s}}},
\end{align}
 where the continuous phonon spectral function, $J_{\rm pn}(\omega) = \alpha_{\rm p} \omega^3 e^{-\frac{\omega^2}{2\omega^2_{\rm p}}}$ and phonon coupling $\alpha_{\rm p} = \frac {D^2}{4\pi^2\hbar c^5_{\rm s} \rho_{\rm d}}$~\cite{Roy3}. The extra factor of $\sqrt{V}$ appearing in (\ref{eq21}) can be absorbed in the higher phonon correlations (e.g. $\avg{\sm b_q}$), when this transformation is used in solving the correlation expansion equations (Sec.~\ref{sec2c}). For polaron and reservoir ME approaches, we use the continuum form of the phonon spectral function $J_{\rm pn}(\omega)$ and with parameters for InAs QDs, with  $\alpha_{\rm p} = 0.06 \,{\rm ps}^2$ and $\omega_{\rm p} = 1$ meV, consistent with experiments~\cite{Weiler}.
The precise value of these phonon parameters will not change
any of the qualitative findings below, and they could be used, e.g., for fitting experiments on a particular QD.

\subsection{Numerical approaches}\label{sec2f}

In the above we have introduced three main theoretical approaches for analyzing the behavior of QDs coupled to structured photonic reservoirs in the presence of phonons, namely polaron MEs (cQED for a simple cavity and bath approaches), correlation expansion and the linear susceptibility technique. The linear susceptibility technique is analytic and computationally the easiest. Both the polaron MEs and the correlation expansion treatments involve simultaneous solutions for first-order coupled ordinary differential equations. The MEs solve coupled equations for the density matrix elements of the QD-photonic reservoir system and the correlation expansion solves coupled Bloch equations for QD-cavity-phonon correlations. The first set of equations (i.e., polaronic ME) is much smaller, since the phonon component is eliminated using the Born-Markov approximation (Sec.~\ref{sec2b}). For the WEA applied here, the relevant basis states for a QD-cavity system are $\ket{01}$, $\ket{10}$ and $\ket{00}$, where in $\ket{XC}$, the first (second) element denotes occupation of exciton (cavity). For this purpose, we use the Quantum Optics (QO) Toolbox~\cite{Tan} to numerically solve the ME, which includes the states $\ket{02}$, $\ket{20}$, $\ket{11}$, for the truncation of the ME~\cite{Tan}. The total density matrix is then expressed in a basis of 6 states. The two-time correlation function $O^{\dagger}(t+\tau)O(t)$ for any operator $O$ can be obtained by solving coupled first-order differential equations for 16 density matrix elements. The 16 elements consist of 6 population elements (diagonal) and 10 polarization elements (first off-diagonal)~\cite{Tan}. The forward matrix propagating the system in time is thus $16 \times 16$ in size. When the polaron reservoir treatment (Sec.~\ref{sec2a}) is used, after elimination of the photonic reservoir, the number of basis states are those of the QD ($\ket{1}$ and $\ket{0}$) and the forward matrix is just $4 \times 4$ in size. In comparison, the non-Markovian correlation expansion treatment scales as $N^2$ in number of correlation elements and $N^2$ x $N^2$ in forward matrix size, where $N$ is the total number of phonon modes used in discretizing the relevant regions of the phonon bands ($\pm 5 $ meV for the current QDs, Sec.~\ref{sec2e}). Finer discretization reduces the numerical noise and for $N$ = 400 (typical numbers), the forward matrix size is $160000 \times 160000$. Thus on a regular work station, the correlation expansion calculations can take a few days  to get the cavity-emitted spectrum compared to polaron ME calculations which take only a few seconds and are this much easier to work with from a numerical perspective.
Moreover, the ME approach can easily include other pump field and nonlinear processes, including multiphoton effects. The correlation expansion is however useful, if one works at low temperature, to justify and test the limits of the polaron ME approaches.

\section{Results}\label{sec3}

\subsection{Emitted spectrum from an uncoupled QD}\label{sec3a}

\begin{figure}[t]
\vspace{0.cm}
\includegraphics[width=0.99\columnwidth]{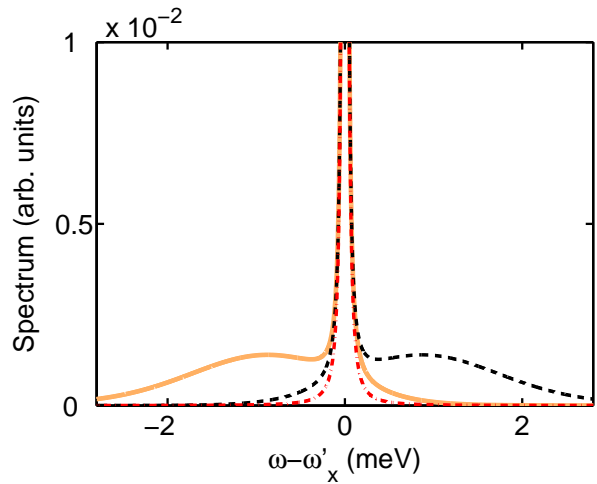}
\vspace{-0.cm}
\caption{(Color online). \footnotesize{ Normalized QD emission (polarization) spectrum (thick orange solid) and linear absorption spectrum (thick black dashed) at T = 4 K for a single uncoupled QD. The red dash-dotted line denotes the polarization spectrum $S^{\rm P}_0$ in the polaron frame. The ZPL parameters are $\gamma_0$ =  5 $\mu$eV, $\gamma_d$ =  5 $\mu$eV.} }
\label{fig3}
\end{figure}

We begin this results section with the  spectrum of a simple QD exciton coupled to phonons, which is decoupled from any structured photonic reservoirs. We expect the spectrum to represent the IBM lineshape due to the presence of phonons with a broadened ZPL, which is a well known result and a useful starting point with which to understand the results when a photon bath is included. In this limit, and at low temperatures, all the models produce qualitatively identical spectra so we only use  the polaron ME model here to analyze the results in this section. The QD is assumed to be weakly excited by an incoherent pump ($P$) and has a background decay rate $\gamma_0$, and pure dephasing rate, $\gamma_d$. In the polaron frame
(Sec.~\ref{sec2b}), the QD emission spectrum (\ref{eq17}) is given by 
\begin{align}
\label{eq21a}
S^{\rm P}_0(\omega) = \lim_{t\rightarrow\infty}\text{Re}[\int_0^{\infty}d\tau\avg{\smdag_{\rm P}(t+\tau)\sm_{\rm P}(t)}e^{i(\omega'_x-\omega)\tau}],
\end{align}
 which is a simple Lorentzian with a total ZPL linewidth $\gamma_0+\gamma_d$,
as shown in  Fig.~\ref{fig3} (dash-dotted  line, without phonon coupling). The transformation from the polaron frame to the lab frame produces a phase term $e^{\phi(\tau)}$ (\ref{eq17a}) and the final QD emission spectrum (i.e., in the lab frame) is 
\begin{align}
\label{eq21b}
S_0(\omega)\!\propto \!\lim_{t\rightarrow\infty}\text{Re}[\int_0^{\infty}
\!d\tau\avg{\smdag_{\rm P}(t+\tau)\sm_{\rm P}(t)}e^{\phi(\tau)}e^{i(\omega'_x-\omega)\tau}].
\end{align}
The resulting emission spectrum (thick orange solid line) is plotted in Fig.~\ref{fig3} and shows the appearance of the phonon sidebands arising due to the IBM phase function $\phi$. The asymmetry of the sidebands arise due to the fact that phonon emission is more probable than absorption at low temperatures, so the phonon emission is more probable on the lower energy side of the ZPL. The Lorentzian ZPL corresponds to the polarization spectrum $S^{\rm P}_0$ in the polaron frame. 

 Similar to Sec.~\ref{sec2d4}, the QD susceptibility function~\cite{Roy3} can be defined as 
\begin{align}
\label{eq21c}
\chi(\omega) \propto i \lim_{t\rightarrow\infty}\int_0^{\infty}d\tau\avg{\sm_{\rm P}(t+\tau)\smdag_{\rm P}(t)}e^{\phi(\tau)}e^{-i(\omega'_x-\omega)\tau},
\end{align}
 and the linear absorption spectrum (Im($\chi$)) is plotted in Fig.~\ref{fig3} (thick black dashed line). The linear absorption spectrum is simply a reflection of the emitted spectrum about the QD ZPL~\cite{Forstner3}. The phonon sidebands are now more enhanced to the right, since phonon
absoprtion is more probably on the higher energy side of the ZPL. To explain this feature, consider a weak coherent laser that excited the QD, i.e., through
 $H_{\rm pump}=\eta_x(\sm e^{i\omega_Lt}+\smdag e^{-i\omega_Lt})$, where $\omega_L$ and $\eta_x$ are the drive strength and the Rabi frequency of the laser, respectively.  A photon at higher frequency ($\omega_L>\omega_x$), can excite the QD more easily by phonon emission process at low temperature. Since the reverse process ($\omega_L<\omega_x$) requires phonon absorption, which is less probable at low temperatures, then absorption is stronger to the right (high energy side) of the ZPL. The linear absorption spectra will be used later in an attempt to understand off-resonant cavity feeding using linear susceptibility theory (Sec.~\ref{sec3c2}).

We next consider coupling to structured reservoirs and begin with the case of a resonant high-Q cavity, i.e.,  in  the strong coupling regime. 

\subsection{On-resonance strong coupling regime in a high-Q cavity}\label{sec3b}

{\bf }

\begin{figure}[t]
\vspace{0.cm}
\includegraphics[width=0.99\columnwidth]{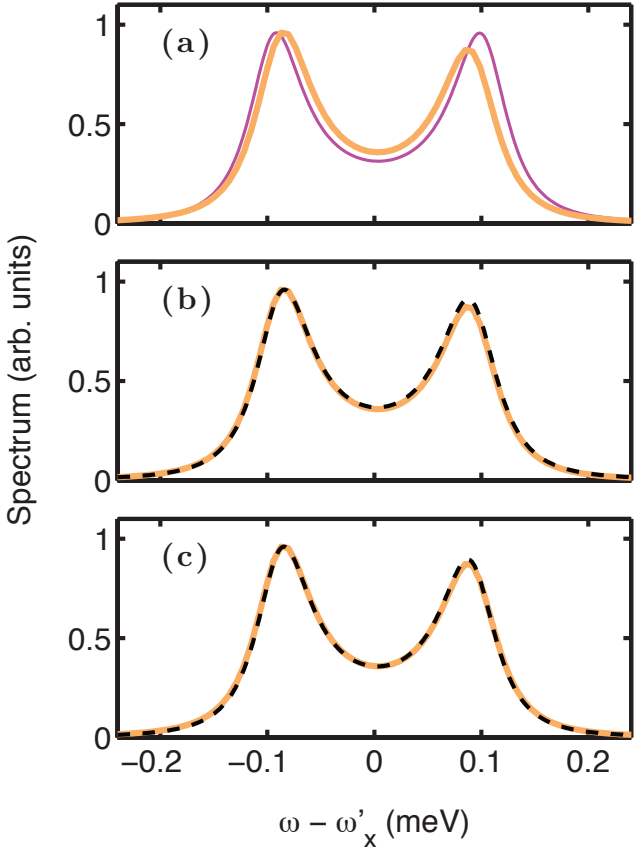}
\vspace{-0.3cm}
\caption{(Color online). \footnotesize{{\bf On-resonance strong coupling regime.} Cavity emission spectra at T = 4 K for a strongly coupled QD-cavity system at resonance ($\omega'_x=\omega_c$). The plots compare coupled mode spectra calculated using different approaches. The thick orange (light solid) line in (a), (b) and (c) denotes the coupled mode spectrum, $S^{\rm CM-ce}_{\rm cav}$ calculated using correlation expansion approach. The magenta (thin solid) and the black (dark) dashed line plots the coupled mode spectrum, $S^{\rm CM-cQED}_{\rm cav}$ calculated using polaron cQED approach, in the absence and presence of phonon coupling in (a)  and (b), respectively. Panel (c) compares the results with linear susceptibility theory and the black (dark) dashed  line denotes the semiclassical Green function spectrum $S^{\rm G-sus}_{\rm cav}$ calculated using a linear susceptibility approach. The main  parameters are $g$ = 100 $\mu$eV, $\kappa$ = 65 $\mu$eV, $\gamma_0$ =  5 $\mu$eV, and  $\gamma_d$ =  55 $\mu$eV.}
 }
\label{fig4}
\end{figure}

As an example of strong coupling between a QD and a photonic reservoir, we consider the case a QD strongly coupled to a high-Q cavity, where the cavity is described by the cavity mode operator $\a$. This is studied
in Fig.~\ref{fig4}. As is well known, a strongly coupled QD-cavity system undergoes vacuum Rabi oscillations, when a single quanta of energy is coherently exchanged between the QD and the cavity~\cite{Yoshie,Bose1}. For a simple two-level atom, coupled to a symmetric cavity without phonon effects, the emission spectrum shows the two hybridized polariton states of equal intensity (magenta, thin solid) line, Fig.~\ref{fig4}(a), separated in frequency by twice the QD-cavity  coupling constant $g$. Such a coherent transfer of energy between light and matter is important for building quantum light-matter interfaces, which could be used, e.g., for long distance quantum communication with photons~\cite{Bose1}. Phonon interactions however affects the coherence of the system. In the presence of phonon coupling, the polaritons appear with different intensities and reduced vacuum Rabi-splitting ($\approx 2\avg{B}g$, orange (thick solid) line, Fig.~\ref{fig4} (a))~\cite{Ota,Milde2}.  Such phonon-dressed strong coupling  can be well explained using the polaron cQED theory (Sec.~\ref{sec2b}), the correlation expansion (Sec.~\ref{sec2c}) or  the linear susceptibility (Sec.~\ref{sec2d1}) approaches, and we compare these different spectra in Fig.~\ref{fig4}. In Fig.~\ref{fig4}(b), we compare the coupled mode spectra derived using the correlation expansion ($S^{\rm CM-ce}_{\rm cav}$ solid line) and cQED polaron ME technique ($S^{\rm CM-cQED}_{\rm cav}$ dark dashed line) and the resultant spectra show very good agreement. The QD ($\omega'_x$) and cavity ($\omega_c$) are assumed to be at a detuning of 0 meV (ZPL) and the frequency of the QD, $\omega'_x$ includes the polaron shift. The parameters used for the simulations are close to typical experimental numbers, which show vacuum Rabi splitting of $2g$ = 200 $\mu$eV~\cite{Ota} ($2g'$ = 183 $\mu$eV) in Fig.~\ref{fig4}. As discussed earlier, the polaron cQED approach is derived using a Born-Markov approximation for the polaron-transformed interaction terms and is valid as long as the system dynamics (i.e., the vacuum Rabi period $\pi/g\approx$ 20 ps (Fig.~\ref{fig4})) is longer than phonon relaxation time ($\approx 3-5$ ps, Sec.~\ref{sec2e}). The correlation expansion approach however uses no such approximations and is expected to work as long as the temperature is sufficiently low (e.g., less than 20 K or so), though it is numerically more cumbersome as discussed earlier (Sec.~\ref{sec2f}). The close agreement (Fig.~\ref{fig4}(b)) of the polaron cQED results (dark dashed line) with the more rigorous non-Markovian correlation expansion results (solid line) demonstrates the validity of polaron cQED ME technique which is also significantly less numerically demanding.

\begin{figure}[t]
\vspace{0.cm}
\includegraphics[width=0.99\columnwidth]{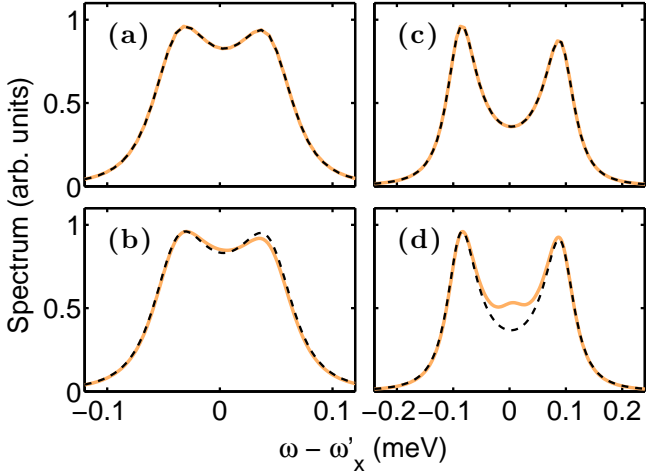}
\vspace{-0.3cm}
\caption{(Color online). \footnotesize{ Normalized emission spectra at T = 4 K for a strongly coupled QD-cavity system at resonance. The plots compare the Green function spectra $S^{\rm G}_{\rm cav}$ (solid line) and the coupled mode spectra $S^{\rm CM}_{\rm cav}$ (dashed line) calculated using correlation expansion (top panels (a, c)) and polaron cQED ME approaches (bottom panels (b, d)). The QD-cavity coupling $g$ is 50 $\mu$eV for (a, b) and 100 $\mu$eV for (c, d). The other parameters are $\kappa$ = 65 $\mu$eV, $\gamma_0$ =  5 $\mu$eV, $\gamma_d$ =  55 $\mu$eV.  } 
}
\label{fig5}
\end{figure}

In Fig.~\ref{fig4}(c), we also compare the coupled mode spectrum obtained from correlation expansion approach $S^{\rm CM-ce}_{\rm cav}$ (solid line) and the semiclassical Green function spectrum obtained using linear susceptibility approach $S^{\rm G-sus}_{\rm cav}$ (dashed line) and they show very good agreement. As described in Sec.~\ref{sec2d4}, the semiclassical linear susceptibility approach  does not make any Markov approximation and evidently works well when the on-resonance condition ($\Delta_{cx'} = 0$)~\cite{HughesSC,savonaPRB} is satisfied. As we will show in the next section, however, 
this semiclassical approach fails to get the correct 
spectra in the off-resonant cavity feeding regime ($\Delta_{cx'}\gg g$). 

As seen above, one of the most striking effects from phonon coupling at the on-resonance ($\Delta_{cx'}=0$) condition is the asymmetric vacuum Rabi doublet~\cite{Ota,Milde2}. The exact form of the ME (\ref{eq9}) however does not bring into light the physical processes responsible for this phonon induced assymetry. The same problem lies with the correlation expansion approach which mathematically reduces to a set of coupled first-order differential equations. The full polaron cQED ME (\ref{eq9}) can however be simplified to an analytical form (\ref{eq9a}), as long as the WEA is valid. This is true for the current situation as we are only dealing with the linear spectrum (as shown in Fig.~(\ref{fig4})). The analytical cQED ME form (\ref{eq9a}) consists of incoherent feeding terms ($\Gamma^{\adag\sm/\smdag\a}$), cross-dephasing term $\gamma_{\rm cd,}$ and  additional terms denoted by $M_1$, $M_2$. For a weakly excited system at resonance, the incoherent feeding terms and the cross-dephasing term $\gamma_{\rm cd}$ have negligible effects on the spectrum. Thus the term solely responsible for the asymmetric doublet is then $M_2$ (as $M_1 =0$ when $\Delta_{cx'}=0$). Such a term  also causes an asymmetry in the ME approaches that describing high-field Mollow triplets~\cite{Ge} when they sample an asymmetric bath. This term then acts as an effective complex coupling $ig'-M_2$ between the resonant QD and cavity (\ref{eq20d}).

The polaron cQED ME approach is strictly valid when the Markov approximation is satisfied and, as was shown above, the coupled mode spectra calculated using the polaron cQED approach matches the more rigorous correlation expansion results closely. It was also mentioned earlier in Sec~\ref{sec2d2}, that the Green function spectrum $S^{\rm G}_{\rm cav}$ and the coupled mode spectrum $S^{\rm CM}_{\rm cav}$ derived using the same theory should match, as long all underlying approximations are valid. In order to test this, we apply non-Markovian correlation expansion and  cQED ME techniques to calculate low temperature (T = 4 K) vacuum Rabi spectrum for parameters which do (Fig.~\ref{fig5} (a, b))~\cite{Bose1} and do not satisfy (Fig.~\ref{fig5} (c, d))~\cite{Ota} the polaronic Born-Markov approximation. 
 The parameters used for both simulations are close to typical experimental numbers, which show vacuum Rabi splitting of $2g$ = 100 $\mu$eV~\cite{Bose1} ($2g'$ = 91.5 $\mu$eV) for Fig.~\ref{fig5} (a, b) and $2g$ = 200 $\mu$eV~\cite{Ota} ($2g'$ = 183 $\mu$eV) for Fig.~\ref{fig5} (c, d). As expected, the Green function spectrum $S^{\rm G-ce}_{\rm cav}$ (thick solid line) and the coupled mode spectrum $S^{\rm CM-ce}_{\rm cav}$ (dark dashed line), calculated using correlation expansion match exactly in both cases (Fig.~\ref{fig5} (a, c)). For Fig.~\ref{fig5}(b), the vacuum Rabi period $\pi/g \approx 40$ ps is much slower than phonon damping (4 ps), which makes the Markov approximation valid. Thus the Green function spectrum $S^{\rm G-cQED}_{\rm cav}$ (solid line) and the coupled mode spectrum $S^{\rm CM-cQED}_{\rm cav}$ (dark-dashed) from the polaron cQED approach match closely. However for Fig.~\ref{fig5}(d), the vacuum Rabi period $\pi/g \approx 20$ ps, and is now  closer to the phonon relaxation time, the Born-Markov
approximation may fail. Hence the coupled-mode spectrum $S^{\rm CM-cQED}_{\rm cav}$ (dark dashed line) now differs strongly from the Green function spectrum $S^{\rm G-cQED}_{\rm cav}$ (thick solid line), which shows an appearance of an unphysical third peak between the two polariton peaks (Fig.~\ref{fig5}(d)). This happens due to the break-down of the Born-Markov approximation, and the polaron cQED approach over-estimates the phonon dressed emission from the QD, at frequencies between the polariton peaks. Hence, when using the polaron cQED theory, the physically correct spectra is obtained using the coupled mode approach. As shown above, it is better
to use the coupled mode approach of the polaron cQED ME
since the cavity emitted spectrum is consistent with the 
input-output theory, while the Green function approach uses
the exact input-output relations and can bring back extra features
that are not contained in the Born-Markov approximation.

It should be noted that a rather large value of pure dephasing $\gamma_d$
 (= 55 $\mu$eV) is used for these spectra calculations. Such a large value is chosen to achieve faster numerical convergence for the correlation expansion calculations, which are numerically intensive. This value of $\gamma_d$ is also used for the off-resonant cavity feeding calculation using correlation expansion,  in the next section.
 The qualitative effects for different values of $\gamma_d$
 remain the same.

\subsection{Off-resonant cavity feeding}\label{sec3c}
\subsubsection{Comparison of three theoretical approaches}\label{sec3c1}

In this section we consider the case of phonon-mediated cavity feeding from an off-resonant QD~\cite{Arka}. Because of the off-resonant condition ($\Delta_{cx'} >$  1 meV), an approximation of weak-coupling between the QD and cavity, is not very restrictive. This allows us to use the polaron reservoir theory (Sec.~\ref{sec2a}) to calculate the spectrum along with the polaron cQED ME approach, correlation expansion and semiclassical linear susceptibility theories. As shown earlier, the reservoir theory accounts for dynamical interplay between photon and phonon baths, which is absent (or restricted to certain regimes) in the other theories~\cite{Kaushik}. 

\begin{figure}[t]
\vspace{0.cm}
\includegraphics[width=0.99\columnwidth]{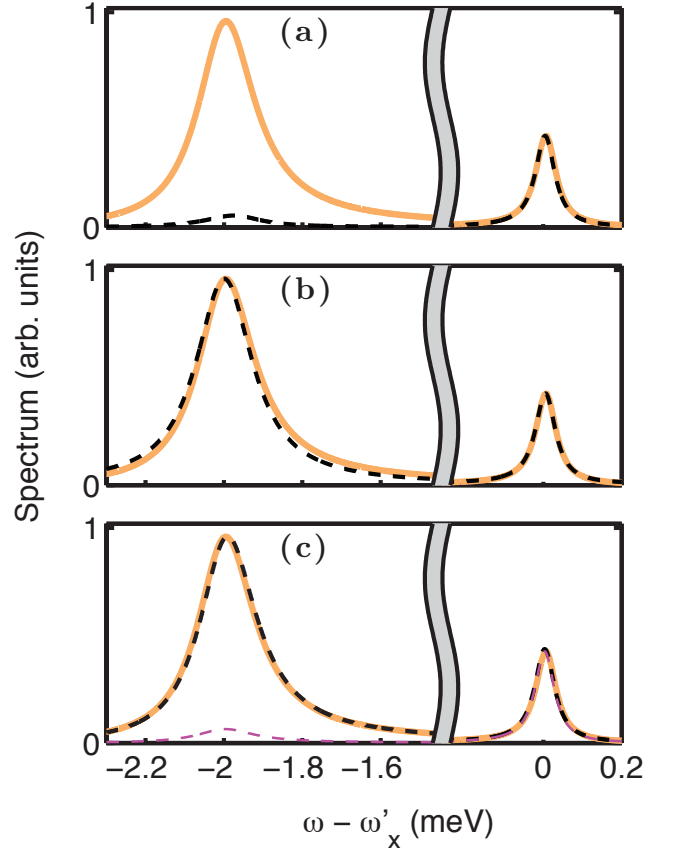}
\vspace{-0.cm}
\caption{(Color online). \footnotesize{{\bf Off-resonance cavity feeding regime.} Normalized emission spectra at T = 4K for a QD weakly coupled to a high-Q cavity (Q = 8000 at $\omega_c/2\pi$ =1440 meV), under off-resonance condition. The thick solid orange (light) line in (a), (b) and (c) denotes the coupled mode spectrum, $S^{\rm CM-ce}_{\rm cav}$ calculated using correlation expansion approach. The black (dark) dashed line plots the coupled mode spectra, $S^{\rm CM-cQED}_{\rm cav}$ in the absence (presence) of phonon coupling in (a) ((b)) using the polaron cQED ME approach.
 Panel (c) compares the correlation expansion calculations (thick solid line) against the Green's function spectrum calculated using polaron reservoir approach ($S^{\rm G-res}_{\rm cav}$, black (dark) dashed) and linear susceptibility ($S^{\rm G-sus}_{\rm cav}$, magenta (light) dashed) approaches, respectively.
  The parameters are $g$ = 100 $\mu$eV, $\kappa$ = 180 $\mu$eV, $\gamma_0$ =  5 $\mu$eV, $\gamma_d$ =  55 $\mu$eV, $\Delta_{cx'}$ = 2 meV. } }
\label{fig6}
\end{figure}

\begin{figure}[ht!]
\vspace{0.cm}
\includegraphics[width=0.99\columnwidth]{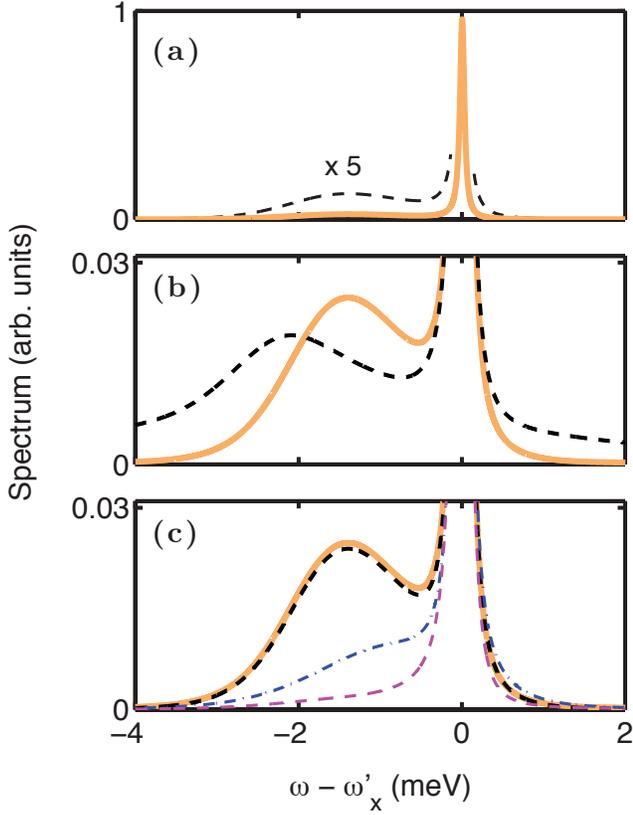}
\vspace{-0.cm}
\caption{(Color online). \footnotesize{ Normalized emission spectra at T = 4 K for a QD weakly coupled to a low-Q off-resonant cavity (Q = 600 at $\omega_c/2\pi$ =1440 meV). The thick solid orange (light) line in (a), (b) and (c) denotes the coupled mode spectrum, $S^{\rm CM-ce}_{\rm cav}$ calculated using correlation expansion approach. The black (dark) dashed line in (a) shows phonon induced feeding and is obtained by magnifying ($\times 5$) the spectrum in this region. Figures (b, c) zoom in on the region near origin, to magnify the phonon effects. Plot (b) compares correlation expansion calculation (solid orange line) with coupled-mode spectra $S^{\rm CM-cQED}_{\rm cav}$ (black (dark) dashed) calculated using polaron cQED ME approach. Plot (c) compares the correlation expansion calculations (solid orange line) against the Green function spectrum calculated using polaron reservoir approach ($S^{\rm G-res}_{\rm cav}$, black  dashed) and linear susceptibility ($S^{\rm G-sus}_{\rm cav}$, magenta (light) dashed) approaches respectively. The blue dash-dotted line shows the polarization spectra $S_0$ calculated using the polaron reservoir approach which resembles the IBM spectra.
 The parameters are $g$ = 100 $\mu$eV, $\kappa$ = 2.4 meV, $\gamma_0$ =  5 $\mu$eV, $\gamma_d$ =  55 $\mu$eV, $\Delta_{cx'}$ = 2 meV.} }
\label{fig7}
\end{figure}

Off-resonant feeding is investigated in the case of a high-Q  (Fig.~\ref{fig6}) and low-Q cavity (Fig.~\ref{fig7}), located approximately - 2 meV away from the QD. In the presence of phonon coupling, the intensity at the off-resonant cavity  is strongly enhanced (Fig.~\ref{fig6}(a), solid line). Such strong feeding is not expected without phonon coupling, when only pure dephasing of the QD is accounted for (Fig.~\ref{fig6}(a), black dashed line). The feeding is much smaller in the low-Q case (Fig.~\ref{fig7}(a), solid line)) and we magnify the region near the origin (Fig.~\ref{fig7}(a), dashed line) to better highlight the cavity feeding effect. The subsequent graphs in the low-Q case (Figs.~\ref{fig7}(b, c)) focus near the origin to investigate off-resonant feeding. 
We compare spectra calculated using the different theoretical approaches in Fig.~\ref{fig6} and Fig.~\ref{fig7}, where 
Note 
the Green function spectrum, $S^{\rm G-ce}_{\rm cav}$ and the coupled mode spectra $S^{\rm CM-ce}_{\rm cav}$ match exactly when calculated using the non-Markovian correlation expansion approach. Hence we use the  coupled mode spectrum, $S^{\rm CM-ce}_{\rm cav}$ (solid orange (thick light) line, Fig.~\ref{fig6} and Fig.~\ref{fig7}), 
to compare against the other approaches. Figures~\ref{fig6} and \ref{fig7} (b) compares the correlation expansion calculations with the polaron cQED ME approach (dashed line). Following the discussion in Sec.\ref{sec3b}, the coupled mode spectrum $S^{\rm CM-cQED}_{\rm cav}$ from the polaron cQED approach is chosen for comparison.  The spectra compare closely in the high Q cavity (Fig.~\ref{fig6} (b)) and show large disagreement in the low Q cavity (Fig.~\ref{fig7} (b)). Figures~\ref{fig6} and \ref{fig7} (c) compares the correlation expansion spectrum with the Green function spectrum derived using the polaron reservoir approach ($S^{\rm G-res}_{\rm cav}$ black (dark) dashed line) and the results show excellent agreement in both high and low Q cases. Thus the coupled mode spectra of the 
polaron cQED approach is not well suited for low Q cavities.

\vspace{-0.01cm}

\begin{figure*}[ht!]
\vspace{-0.2cm}
\includegraphics[width=0.77\textwidth]{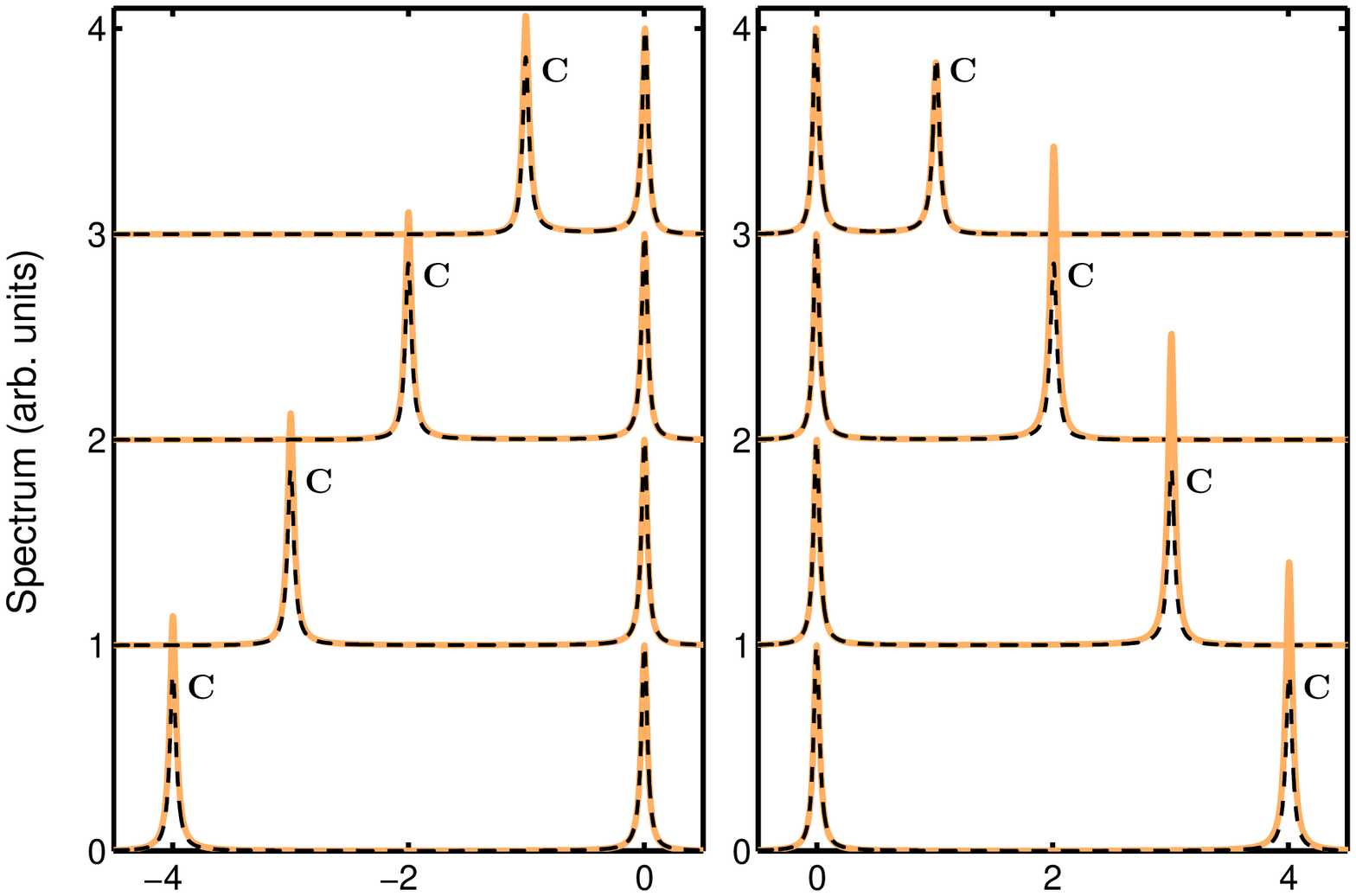}
\vspace{-0.85cm}
\includegraphics[width=0.77\textwidth]{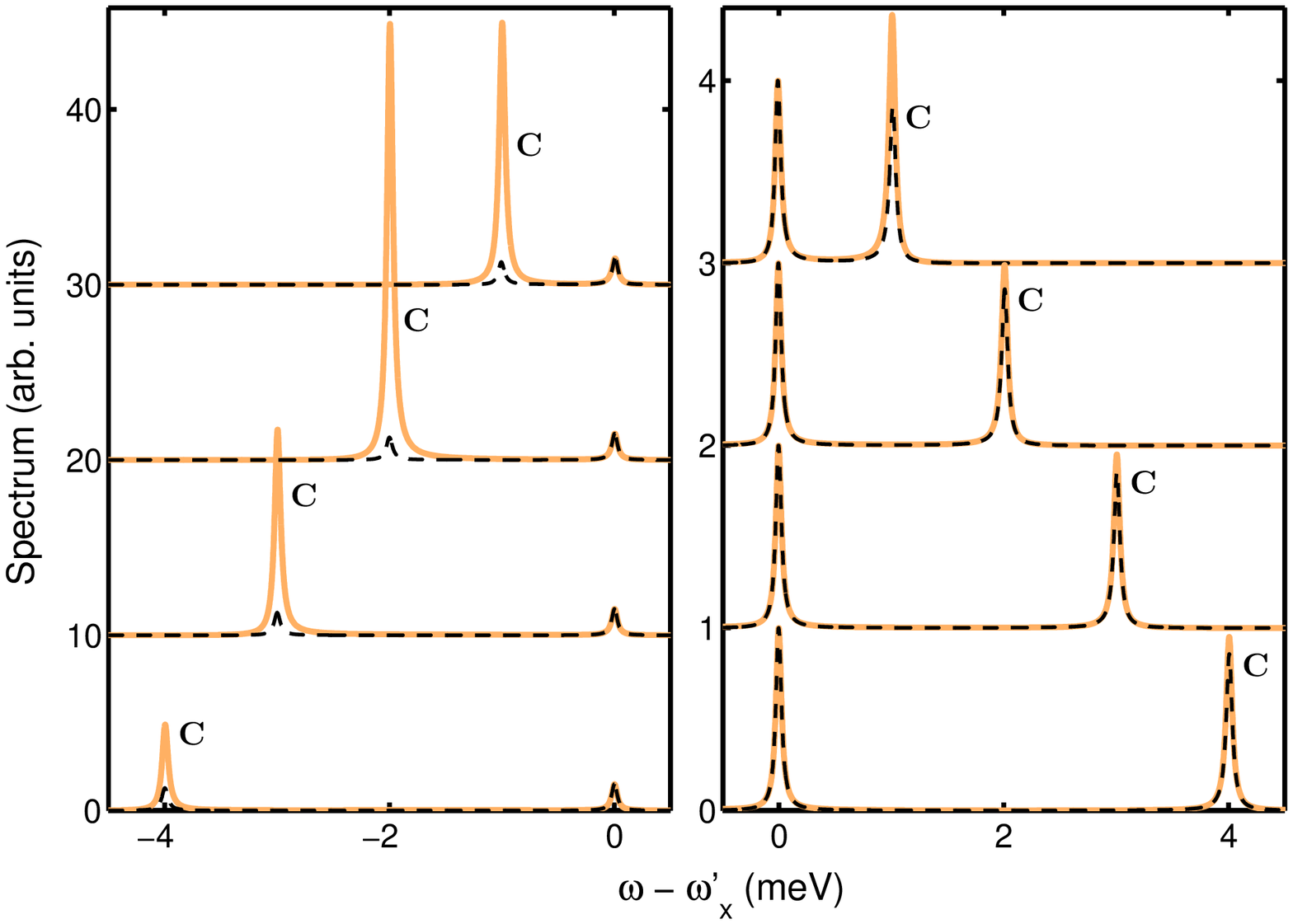}
\vspace{-0.cm}
\caption{(Color online). \footnotesize{{\bf Off-resonance cavity feeding regime: polaron ME versus linear susceptibility model.} Normalized Green function spectra for a QD-cavity system without phonons (black dashed) and with phonons (T = 4 K) (solid orange) calculated from linear susceptibility method $S^{\rm G-sus}_{\rm cav}$ (top panels) and the polaron reservoir method, $S^{\rm G-res}_{\rm cav}$(bottom panels). The QD-cavity detunings are 1, 2, 3, 4 meV and -1, -2, -3, -4 meV (top to bottom) in left and right panels respectively. Individual spectra are normalized by peak ZPL intensity and data in bottom-left panel is multiplied by 1.5 for better visibility. The main parameters are $g$  = 100 $\mu$eV, $\kappa$ = 65 $\mu$eV, $\gamma_0$ =  5 $\mu$eV, $\gamma_d$ =  55 $\mu$eV.} }
\label{fig8}
\end{figure*}

The emission spectrum (solid orange line) resembles a sum of two coupled Lorentzians in the high Q cavity (Fig.~\ref{fig6}(c)). In the low Q cavity however, the emission spectra has a non-Lorentzian shape (Fig.~\ref{fig7}(c), solid orange line) which resembles the IBM spectral lineshape. Physically, for low Q cavities the emission spectrum should resemble the non-Lorentzian IBM spectrum as demonstrated experimentally by Weiler {\it et  al.}~\cite{Weiler}, where low Q cavities are used to measure vertical emission from QDs. The Green function spectrum from the reservoir theory $S^{\rm G-res}_{\rm cav}$ correctly reproduces the spectrum (black (dark) dashed line Fig.~\ref{fig7}(c)) in the high and low Q case and the reason for this can be explained by understanding the way in which it is calculated. The calculation of the Green function spectrum using the reservoir approach uses the polarization spectra $S_0$ (\ref{eq21b}) followed by subsequent projection using the cavity propagator $\alpha^{\rm cav}_{\rm prop}$ (\ref{eq18}),
which is an exact input-output formalism. The SE rate of the QD is given by $\tilde{\gamma}$~\cite{Kaushik} using the photon reservoir approach (\ref{eq5}). The exciton polarization spectra in the polaron frame (\ref{eq21a}), $S^P_0(\omega) = \lim_{t\rightarrow\infty}\text{Re}[\int_0^{\infty}d\tau\avg{\smdag_{\rm P}(t+\tau)\sm_{\rm P}(t)}]$ 
 is thus a Lorentzian (dark dash-dotted line, Fig.~\ref{fig3})  with spectral width determined by total dephasing ${\tilde \gamma}+\gamma_0+\gamma_d+P$, where the latter is negligible here. This Lorentzian (dash-dotted line, Fig.~\ref{fig3}) is the  ZPL of the QD. The non-Lorentzian sidebands (thick light solid line, Fig.~\ref{fig3}) of the QD emission spectrum in the lab frame $S_0$ (\ref{eq21b}) due to acoustic phonon interaction~\cite{Weiler}  arise as the phonon correlation function $e^{\phi(\tau)}$ is analytically included in the QD correlation (\ref{eq17a}) when a transformation is made from polaron ($S^P_0$) to lab frame ($S_0$) (see discussion in Sec.~\ref{sec3a}). 
This gives $S_0$ the characteristic non-Lorentzian lineshape of the IBM (Fig.~\ref{fig3}, light solid line) and is plotted for the low Q case in Fig.~\ref{fig7}(c) (blue dashed-dotted line)~\cite{Besombes}. The final Green function spectrum $S^{\rm G-res}_{\rm cav}$ (Fig.~\ref{fig7}(c), black (dark) dashed line) is obtained by multiplying the polarization spectrum $S_0$ (Fig.~\ref{fig7}(c), blue dashed-dotted line), with the cavity propagator $\alpha^{\rm cav}_{\rm prop}$ (\ref{eq18}), which amplifies the phonon sidebands. In the high Q case, the cavity propagator amplifies a narrow region of the phonon side-band around $\omega_c$ and the Green function spectrum from the reservoir approach $S^{\rm G-res}_{\rm cav}$ resembles a sum of two Lorentzians at QD and cavity. This is however not the situation in the low Q cavity (Fig.~\ref{fig7}(c)). For a low-Q cavity, the position of the second peak due to cavity amplification strongly depends on the phonon sideband spectrum and may appear at a position different from its original location at $\omega_c\approx -2$ meV~\cite{Valente}. The process also makes the resultant spectra $S^{\rm G-res}_{\rm cav}$ (black (dark) dashed line) in Fig.~\ref{fig7}(c) resemble the IBM spectrum ($S_0$, blue dash-dotted, Fig.~\ref{fig7}(c)) more closely. Thus the projected spectra from the photon reservoir theory matches the physically correct results closely and provides a huge computational advantage over the more numerically demanding correlation expansion technique. 

The coupled mode spectrum $S^{\rm CM-cQED}_{\rm cav}$ from the polaron cQED approach resembles a sum of two Lorentzian modes with the QD and the cavity peaks appearing at their original positions $\omega'_x = 0$ and $\omega_c\approx$ 2 meV, respectively (Fig.~\ref{fig6}(b) and Fig.~\ref{fig7}(b), black dashed line). This is close to the correct spectral shape in the high Q case (solid orange line, Fig.~\ref{fig6}(b)) but deviates strongly from it in the low Q case. In the low Q case, the non-Markovian dynamics is not correctly accounted in polaron cQED coupled-mode spectra $S^{\rm CM-cQED}_{\rm cav} \propto \avg{\adag(\omega)\a(\omega)}$, which always resembles a sum of two Lorentzians. This non-Markovian dynamical effect is captured by the correlation expansion calculation as well as the Green function spectrum from the reservoir approach. In the reservoir approach, this dynamics is accounted in the QD polarization spectra $S_0$, which is subsequently multiplied by the Lorentzian cavity projector to calculate $S^{\rm G-res}_{\rm cav}$.

A slight disagreement between the projected reservoir spectrum $S^{\rm G-res}_{\rm cav}$ (black  dashed) and coupled mode correlation expansion spectrum $S^{\rm CM-ce}_{\rm cav}$ (orange solid) can be observed in the low Q case (Fig.~\ref{fig7}(c)) as compared to the high Q case (Fig.~\ref{fig6}(c)).
This difference arises, because the reservoir theory accounts for the dynamical interplay between the photon and phonon bath in determining the SE rate $\tilde{\gamma}$ (\ref{eq5}); $\tilde{\gamma}$ in turn determines the spectral-width of the ZPL in the QD polarization spectra $S_0$ (\ref{eq21b}). The structure of the photonic reservoir is more correctly incorporated in the reservoir approach as compared to correlation expansion technique, which incorporates the photon bath at the level of a system operator $\a$ and accounts for its structure / damping using phenomenological decay term $\kappa$. In the latter case $\tilde \gamma =\tilde \gamma_{\rm P} =  \Gamma^{a^\dagger\sigma^-}_0+ 2g'^2\frac{\frac{\kappa}{2}}{\Delta_{cx'}^2+({\frac{\kappa}{2}})^2} $~\cite{Kaushik} (\ref{eq9e}). Such an approximation is strictly valid for a high Q cavity, where the spectral width of the photon bath is negligible compared the phonon reservoir band-width (8-10 meV) and breaks down ($\tilde \gamma \neq \tilde \gamma_{\rm P}$) when both reservoirs have comparable spectral widths. This is the case for the low Q cavity. The difference is however very small here because the net dephasing rate ${\tilde \gamma}+\gamma_0+\gamma_d+P$ determining the spectral width of the Lorentzian ZPL of $S_0$ is dominated by the pure dephasing ($\gamma_d$ = 55 $\mu$eV) in the current case, which is much larger than $\tilde \gamma \approx$ 1-4 $\mu$eV. 

A comparison between the spectra from the correlation expansion approach $S^{\rm CM-ce}_{\rm cav}$ (light solid), and linear susceptibility theory $S^{\rm G-sus}_{\rm cav}$ (magenta (thin light) dashed) is also shown in Fig.~\ref{fig6}(c) and \ref{fig7}(c), where the linear susceptibility theory shows very little feeding to the off-resonant cavity. This discrepancy with linear susceptibility theory in the context of off-resonant feeding was mentioned earlier in Sec.~\ref{sec3b} and is explained in detail in Sec.~\ref{sec3c2} below.

\vspace{-0.01cm}


\subsubsection{Explanation of the failure of the susceptibility theory for explaining the   off-resonant cavity feeding}\label{sec3c2}

In this section we compare the Green function spectrum calculated using the polaron reservoir ME (\ref{eq4}) and the linear suspecpibility approach to obtaining the spectrum (\ref{eq20h}), for the case of off-resonant cavity feeding and explain the large differences between the two results shown previously in Fig.~\ref{fig6} and \ref{fig7}(c). As shown in Ref.~\onlinecite{HughesSC}, the cavity-emitted spectrum $S^{\rm G-sus}_{\rm cav}$ (i.e. \ref{eq20h}) is obtained by multiplying the linear susceptibility $\chi$ with a Lorentzian cavity projector (\ref{eq18}). The Im($\chi$) (linear absorption spectra) plotted in Fig.\ref{fig3} (dark dashed line) shows that the phonon induced enhancements (asymmetries) are stronger to the right, than to the left of the ZPL. The same holds true for the Re($\chi$) (not shown). Thus when multiplied by a Lorentzian projector for the spectrum (\ref{eq20h}), cavity feeding is stronger to the right than to the left~\cite{ HughesSC, savonaPRB}. This is seen from the graphs in the top panel of Fig.~\ref{fig8}, where normalized spectra (light solid orange line) are plotted for different detunings in the presence of phonons. The dark dashed lines represent the corresponding normalized spectra in the absence of phonons. Individual spectra are normalized with respect to the peak ZPL intensity. Left (right) panels plots spectra for positive (negative) QD-cavity detunings ($\Delta_{x'c}$). Cavity feeding can be estimated from the ratio of peak heights as both Lorentzians have comparable linewidths ($\Gamma_x = 60\, \mu$eV and $\kappa = 65\, \mu$eV) and it increases in presence of phonons. The feeding is asymmetric and stronger when the cavity is to the right. These results are however in complete contrast with the polaron reservoir ME calculations (bottom panels, solid orange lines), which shows a stronger cavity enhancement to the left of the QD ZPL. The dark dashed lines once again plot the normalized spectra without phonons. As explained earlier (Sec.~\ref{sec2d1}), the Green function spectra, $S^{\rm G-res}_{\rm cav}$ (\ref{eq16}) is obtained by multiplying the polarization spectrum $S_0$ of the QD (Fig.~\ref{fig3}, solid line), with the Lorentzian cavity projector (\ref{eq18}). The QD polarization spectrum (Fig.~\ref{fig3}, solid line) shows strong phonon bands to the left of the QD ZPL and hence the cavity is fed more strongly to the left at low temperatures. 

 We have so far provided a mathematical explanation for this difference in cavity feeding. The results from the polaron reservoir ME can however be justified to be physically correct. At low temperatures phonon emission is more probable than phonon absorption. Thus if a cavity has lower energy than an excited QD, the cavity can be excited more easily in a two-step quantum process where a cavity photon is created along with a phonon emission. When $\omega_c > \omega'_x$, exciting the cavity will require phonon absorption which is less feasible at low temperatures. Thus the polaron reservoir ME predicts the physically correct solution. The linear susceptibility theory misses this two step quantum process which is correctly incorporated in the SE rate (\ref{eq9e}) calculated using polaron reservoir ME approaches, through the Lindblad QD/cavity scattering rates $\Gamma^{\smdag \a/\adag \sm}$. The linear susceptibility technique is however very successful in explaining the on-resonance ($\Delta_{cx'} = 0$) phonon induced assymetric vaccum Rabi doublet . This is because at resonance, these two-step quantum feeding processes barely contribute to the emission spectra (see Sec.~\ref{sec3b}). The technique however fails for far detuned condition ($\abs{\Delta_{cx'}} \gg 0$), where these two-step quantum processes become important.

\subsection{Emission spectra from a slow-light coupled-cavity waveguide}\label{sec3d}

\begin{figure}[t!]
\vspace{0.cm}
\includegraphics[width=0.99\columnwidth]{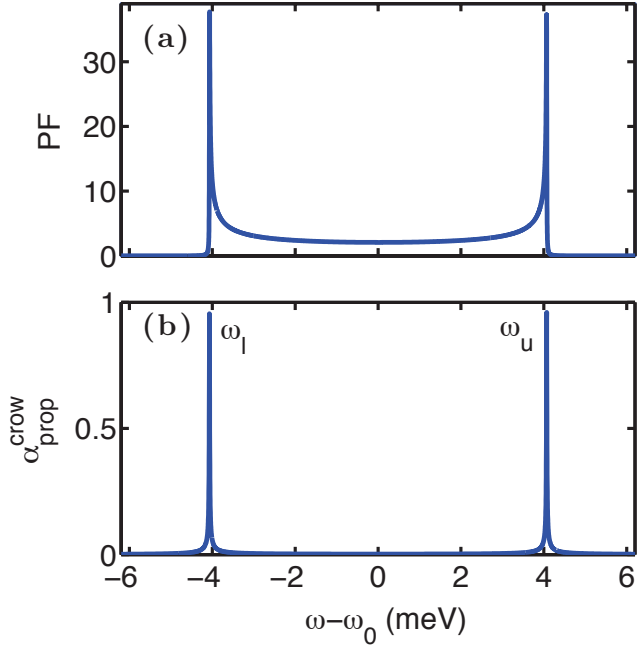}
\vspace{-0.cm}
        \caption{(Color online). {\bf Slow-light CROW Purcell factor and propagator.} \footnotesize{ (a) Purcell factor (PF) and (b) projector function $\alpha_{\rm P}$ for a coupled-cavity photonic crystal waveguide (CROW). The parameters used for waveguide calculations are taken from Ref.~\onlinecite{Kuramochi} (see text)  and QD-waveguide coupling $g = \left(\frac{d^2 \omega_0}{2\hbar \epsilon_0\epsilon V_{\rm eff}}\right )^{\frac{1}{2}}$ = 85 $\mu$eV (with $\mu=50$ Debye), where $\omega_0$ marks the waveguide band-center.} }
\label{fig9}
\end{figure}

The reservoir theory can be applied to general structured reservoirs, beyond simple Lorentzian cavities. In photonic structures with a non-Lorentzian LDOS profile, the dynamical interplay between photon and phonon reservoir can manifest itself more strongly, as has been demonstrated by photoluminescence intensity studies in coupled-cavity waveguides~\cite{Kaushik2}. In this section we calcuate the emission spectra $S^{\rm G-res}_{\rm wg}$ of a QD coupled to a photonic crystal waveguide, using the reservoir theory. Photonic crystal waveguides are important for slow light applications~\cite{Vlasov, Baba, Notomi2} and on chip single photon emission~\cite{Hughes1, Rao1, Laucht, Shields}. The specific case of a photonic crystal coupled-cavity waveguide (CROW)~\cite{Notomi2} is considered here  for which a simple expression of the medium electric field Green function, $\v{G}$ can be derived analytically, using a tight-binding approach~\cite{Yariv, Fussell1, Fussell2}. The Green function is in turn used to determine the LDOS and the projector $\alpha_{\rm P}$ (\ref{eq19}) of the wave-guide. The Purcell factor (PF) and the projector is plotted in Fig.~\ref{fig9}(a) and (b), respectively. The PF (which depends on the projected LDOS) is given by $\gamma/\gamma_b$, where $\gamma = 2\int_0^{\infty}\text{Re}[J_{\text{ph}}(\tau)]d\tau$~\cite{Kaushik} is the frequency-dependent SE rate of the QD in the waveguide without phonon coupling, and $J_{\text{ph}}(\tau)$ is the waveguide bath relaxation function (Sec.~\ref{sec2a}) and $\gamma_b = d^2\sqrt{\epsilon}\,\omega^3/(6\pi\hbar\epsilon_0 c^3)$ is the decay rate in the background slab material. As shown in Fig.~\ref{fig9}(b), the projector can be approximated as a sum of two sharp Lorentzians located at the upper ($\omega_u$) and lower ($\omega_l$) mode-edges of the waveguide. The parameters used correspond to a CROW structure built by coupling individual cavities formed by local width modulation of a line defect waveguide~\cite{Kuramochi}, on a photonic crystal slab, which has  a period of $d = 420$ nm ~\cite{Fussell2}. The individual cavities have a 
mode volume of $V_{\rm eff}$ = 0.175 x $10^{-18}$ ${\rm m}^{3}$ and the CROW period is $\approx 5d$ ~\cite{Fussell2,Notomi1}. 

\begin{figure}[ht!]
\vspace{0.cm}
\includegraphics[width=0.99\columnwidth]{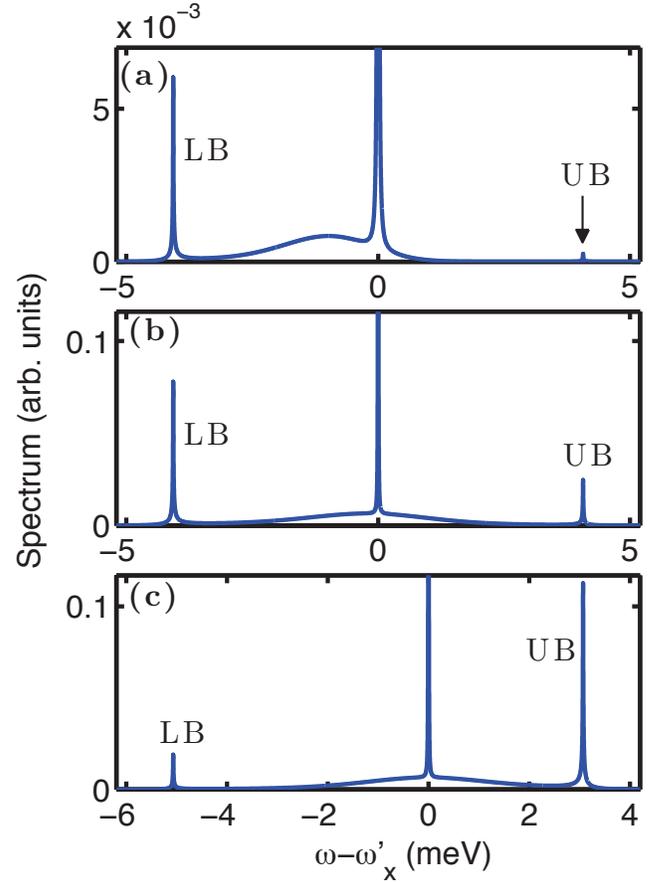}
\vspace{-0.cm}
        \caption{(Color online).  \footnotesize{ Normalized emission spectra $S^{\rm G-res}_{\rm wg}$ for a QD weakly coupled to a waveguide. (a) and (b) plot the spectrum at T = 4 and 40 K, respectively, when $\omega'_x = \omega_0$, where $\omega_0$ marks the waveguide band-center and (c) is the spectrum at T = 40 K and $\omega'_x-\omega_0$ = 1 meV. The UB and LB mark the upper and lower waveguide bands. The parameters are $g$ = 85 $\mu$eV, $\gamma_0$ = $\gamma_d$ =  1 $\mu$eV.} }
\label{fig10}
\end{figure}

The polarization spectrum from a bare QD resembles the IBM spectrum~\cite{Besombes, Weiler} (Fig.~\ref{fig3}, solid line). In the weak-coupling limit, the spectrum of a QD coupled to waveguide (Fig.~\ref{fig10}) is obtained by multiplying the IBM spectrum with the projector (\ref{eq19a}). Figures~\ref{fig10} (a) and (b) plot the waveguide spectrum $S^{\rm G-res}_{\rm wg}$ (normalized units) at T = 4 and 40 K respectively, for a QD located at the band-center, $\omega_0$ (= $\omega'_x$). As shown in Fig.~\ref{fig10}(b), the projector amplifies the phonon side-bands and leads to the appearance of three distinct emission peaks, which corresponds to the two waveguide mode-edges and the QD ZPL. The peak intensity of the three peaks can be denoted as  $I_0 = S^{\rm G-res}_{\rm wg}(\omega'_x)$ (QD-ZPL peak), $U_0 = S^{\rm G-res}_{\rm wg}(\omega_u)$ (upper waveguide band (UB) at $\omega =\omega_u$) and $L_0 = S^{\rm G-res}_{\rm wg}(\omega_l)$ (lower waveguide band LB at $\omega =\omega_l$). The emission at the lower mode-edge ($L_0$) is stronger than the upper mode-edge ($U_0$), primarily because of the asymmetry of the phonon sidebands due to a stronger phonon emission probability at low temperatures~\cite{Weiler}. The phonon-assisted ``feeding'' of the waveguide mode-edges can be varied by tuning the QD in frequency as shown by Fig.~\ref{fig10}(c), which plots $S^{\rm G-res}_{\rm wg}$ (at T = 40 K) for a QD located 1 meV to the right of the band-center ($\omega_0$). As seen from the normalized spectra of Fig.~\ref{fig10} (a) and (b), mode-edge feeding due to phonons become stronger at higher temperatures. Hence we use a higher temperature (T = 40 K) for further investigating phonon effects below. For the moment, any temperature dependence of pure dephasing is neglected (Fig.~\ref{fig10}), but we will include this effect below (Fig~\ref{fig13}(b)).

\begin{figure}[ht!]
\vspace{0.cm}
\includegraphics[width=0.99\columnwidth]{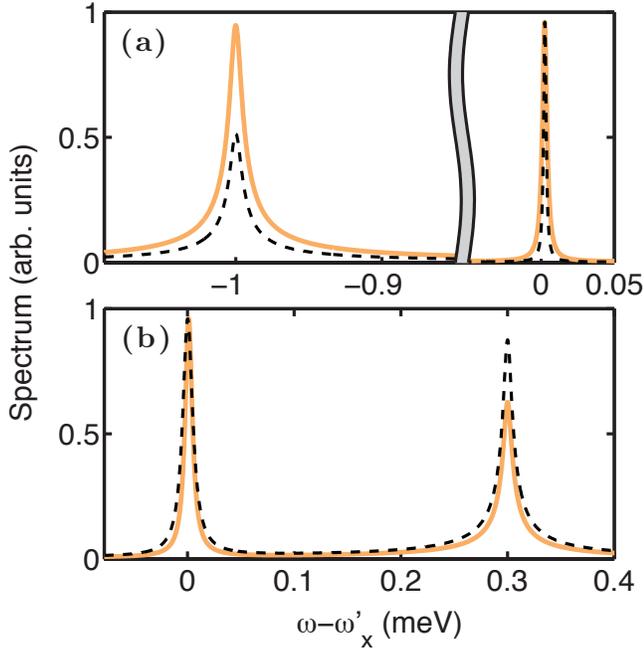}
\vspace{0.cm}
        \caption{(Color online). \footnotesize{ Normalized projected spectra $S^{\rm G-res}_{\rm wg}$ at T = 40 K in a CROW, for a QD (at 0 meV) located 1 meV to the right (a) and 0.3 meV to the left of the left of the upper modeedge (b). The orange (light) solid (black (dark) dashed) line plots spectra in case when phonons do ( do not ) influence the SE rate, $\tilde\gamma$ ($\gamma$). The parameters are $g$ = 85 $\mu$eV, $\gamma_0=\gamma_d$ =  1.0 $\mu$eV.} }
\label{fig11}
\end{figure}

\begin{figure}[t!]
\vspace{0.cm}
\includegraphics[width=0.99\columnwidth]{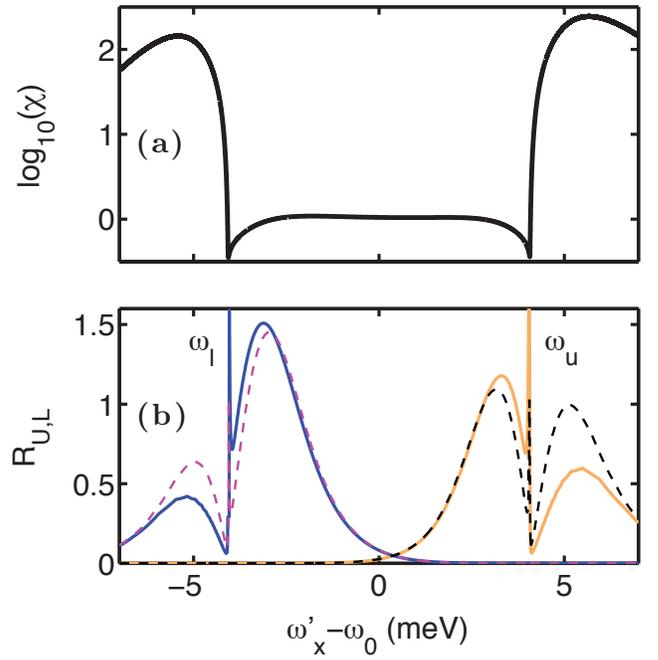}
\vspace{-0.cm}
        \caption{(Color online). \footnotesize{(a) Spontaneous emission enhancement factor $\chi$ (at T = 40 K) plotted in log10 scale, as the frequency of the QD is scanned across the waveguide LDOS. (b) Lower and upper branch peak ratios plotted as the QD frequency is scanned across the waveguide LDOS at T = 40 K. The blue (dark) and orange (light) solid line plots lower and upper branch ratios respectively, when phonons do not influence SE rates ($\gamma$). The magenta (light) and black (dark) dashed lines plot lower and upper branch ratios respectively, when phonons do influence SE rates ($\tilde\gamma$). $\omega_0$ marks the band-center frequency of the waveguide. The parameters are $g$ = 85 $\mu$eV, $\gamma_0 = \gamma_d$ =  1 $\mu$eV.} }
\label{fig12}
\end{figure}

\begin{figure}[t]
\vspace{0.cm}
\includegraphics[width=0.99\columnwidth]{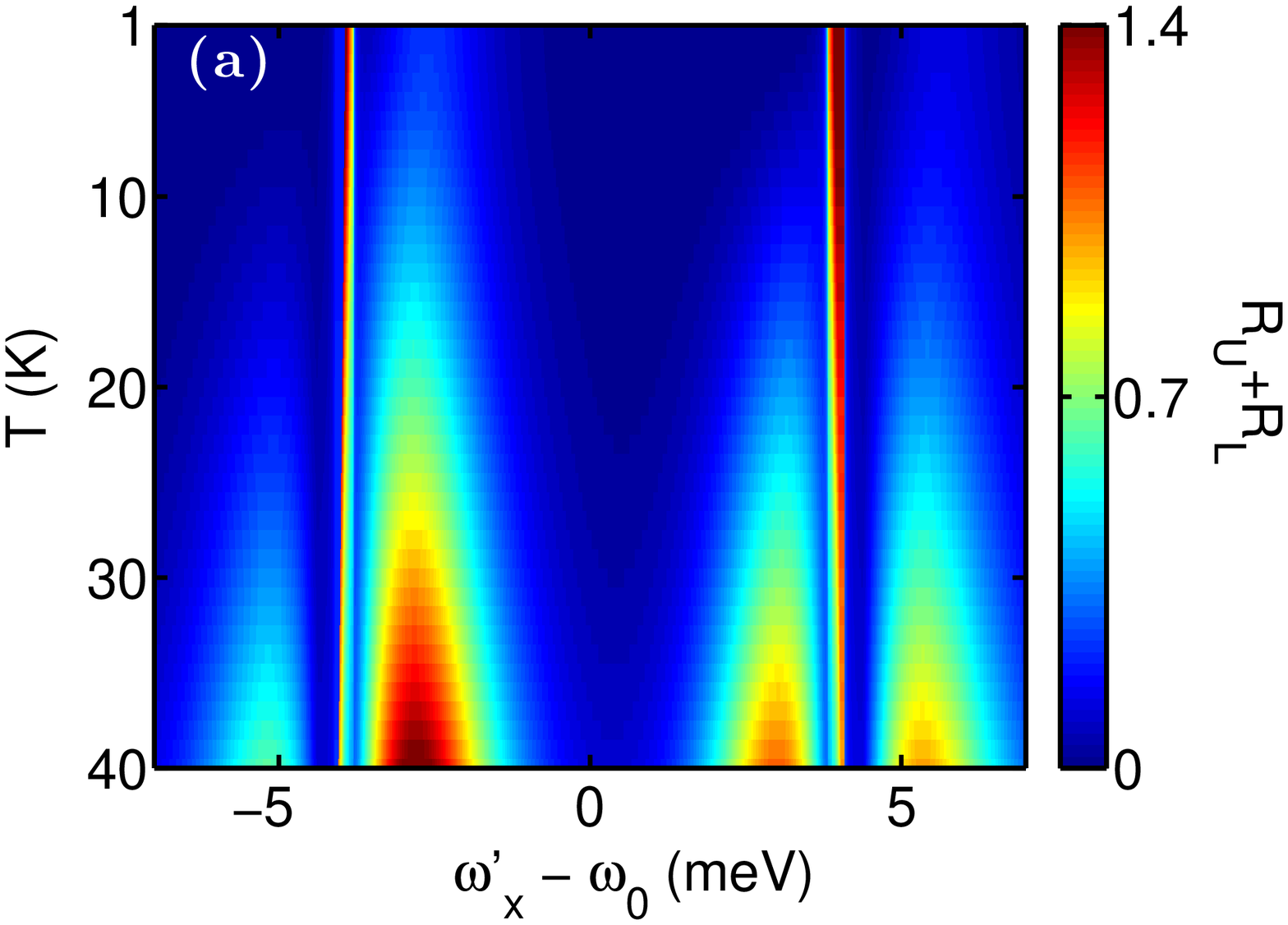}
\vspace{-0.cm}
\includegraphics[width=0.99\columnwidth]{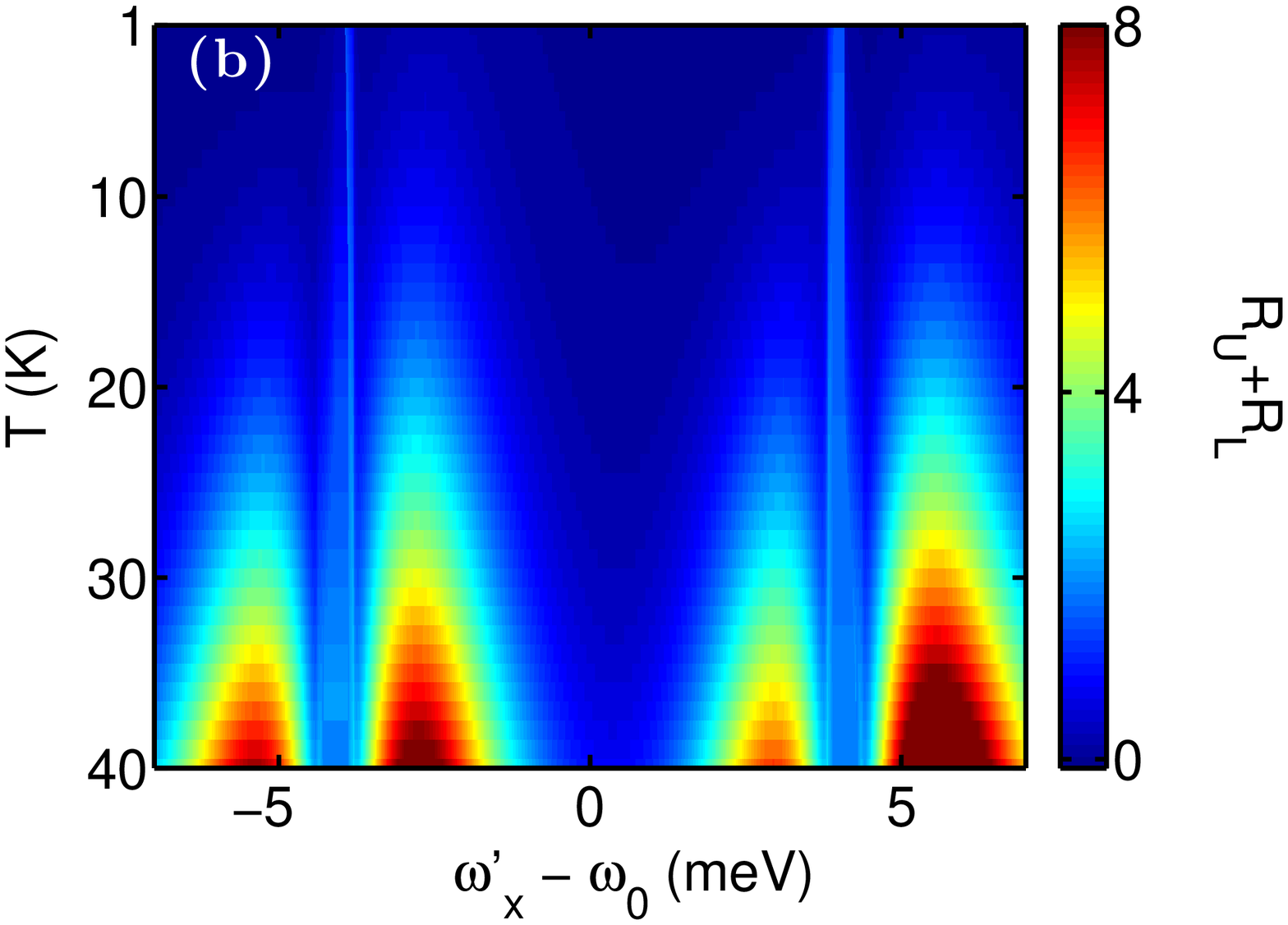}
\vspace{-0.cm}
        \caption{(Color online). \footnotesize{ A sum of the feeding ratio $R_U+R_L$
         plotted as a function of temperature and detuning. The parameters are $g$ = 85 $\mu$eV, $\gamma_0$ =  1 $\mu$eV and $\gamma_d$ = 1 $\mu$eV for (a) and $\gamma_d$ = 1 +0.95(T-1) $\mu$eV for (b). The color scale in (b) is saturated at a ratio of 8 to show features at lower temperatures.} }
\label{fig13}
\end{figure}

As also observed from Fig.~\ref{fig10},  the spectral width of the waveguide (8 meV) is comparable to the phonon reservoir. The dynamical interplay between the two reservoirs is expected to influence the emission spectra in this case. As shown elsewhere~\cite{Kaushik, Kaushik2}, this interaction leads to a strong enhancement of SE rates outside the waveguide band (also Fig.~\ref{fig12}(a)), where SE otherwise is strongly suppressed (Fig.~\ref{fig9}(a)). For the case of the waveguide  emission spectrum, the effect of this interplay can be observed by tuning the QD close to the upper mode-edge (Fig.~\ref{fig11}). Since the lower mode-edge is located at a frequency larger than the spectral width of the phonon side-bands, its emission intensity is negligible and hence not included in the current figure. In Fig.~\ref{fig11} (a), the QD (at 0 meV) is tuned 1 meV outside the upper mode-edge. The upper mode-edge appears clearly on the spectrum due to the presence of the phonon sidebands. The solid line plots the emission spectrum, when the SE rate ($\tilde\gamma$) of the QD is estimated by taking the effects phonon bath into account (\ref{eq5}). The dashed line plots the spectra, for the case when phonons do not influence the SE rate ($\gamma=2\int_0^{\infty}\text{Re}[J_{\text{ph}}(\tau)]d\tau$)~\cite{Kaushik}. Due to the phonon-mediated SE enhancement outside the wave-guide mode edge, the solid line shows a stronger feeding of the waveguide compared to dashed line. The reverse effect happens in Fig.~\ref{fig11} (b) when the QD is tuned inside the waveguide band by about 0.3 meV to the left of the upper mode-edge. Here phonon modification leads to reduction of SE rate, $\tilde\gamma<\gamma$~\cite{Kaushik} (also Fig.~\ref{fig12}(a)). Thus when phonons influence the SE rate, the feeding of the waveguide mode-edge (solid line) is lower. 

The extent of feeding of the upper and the lower waveguide mode-edge by the QD can be quantified using peak intensity ratios as upper branch ratio, $R_U  = U_0/I_0$ and lower branch ratio, $R_L  = L_0/I_0$. These intensity ratios is then plotted (Fig.~\ref{fig12}(b)) as the QD is scanned in frequency across the waveguide LDOS at T = 40 K (see Fig.~\ref{fig10} (b, c)). The blue (dark) and orange (light) solid lines plot the lower and upper branch ratios respectively, when phonons do not influence the SE rate ($\gamma$).  The magenta (light) and black (dark) dashed lines plot the lower and upper branch ratios respectively, when phonons influence the SE rate, ($\tilde\gamma$). The SE modification factor $\chi = \tilde\gamma/\gamma$~\cite{Kaushik} in presence of phonons as a function of wavguide LDOS frequency is also reproduced for reference in Fig.~\ref{fig12}(a). The broad side-bands in the peak ratios about the waveguide mode-edges appear due to intensity enhancement due to phonon side-bands. The sharp peak in the ratio happens near the mode-edge as the QD ZPL is scanned across the sharp waveguide mode-edge (width = 14 $\mu$eV). The lower branch ratio $R_L$ is stronger and wider than the upper branch ratio $R_U$ inside the waveguide band, because lower (upper) branch appears due to phonon emission (absorption). For the same reasons, the upper branch ratio $R_U$ is larger and broader than the lower branch ratio $R_L$, outside the waveguide band. In the case when phonons do not influence SE ($\gamma$), as the QD is tuned across the wave-guide, stronger feeding is expected when the QD is aligned to the right of a mode-edge, due to stronger phonon emission~\cite{Weiler}. This is true for the lower branch ratio $R_L$ without phonons (blue (dark) solid) which is stronger inside the waveguide band compared to outside. The reverse happens though for the upper branch ratio $R_U$ without phonons (orange (light) solid).  In case of the upper mode-edge, the reverse happens, because the overall SE rate ($\gamma$, Fig.~\ref{fig9}(a)) of the QD reduces as it moves outside the waveguide band. This effect is somehow suppressed when the phonon modification of SE rate ($\gamma \rightarrow \tilde\gamma$, Fig.~\ref{fig12}(a)) is accounted for (black (dark) dashed). The feeding ratio is substantially larger in this case (almost twice), compared to the case where such effects are ignored ($\gamma$). Phonons have an opposite effect however at and near the mode-edge inside the waveguide band. Here the feeding is suppressed (dashed) due phonon induced reduction of SE rate~\cite{Kaushik}. 

A better understanding of the mode-edge feeding can be obtained by plotting the ratios as a function of temperature and detuning (Fig.~\ref{fig13}(a)). Due to very small spectral overlap between the ratios (Fig.~\ref{fig12} (b)), a sum of the two ratios $R_U+R_L$ is plotted. The intensity of the sidebands increase with respect to the mode-edge peak with temperature, as the phonon sidebands become stronger.  At very low temperatures, when phonon absorption is negligible~\cite{Weiler}, the feeding is stronger as expected, when the QD is aligned to the right of mode-edge. The trend however reverses for the upper mode-edge as T is increased (\ref{fig12}(b), dark dashed lines). This is because the total SE rate inside the upper mode-edge is always greater than outside for the highest temperatures considered in the plot.

The feeding ratios increase almost by a factor of 10 (Fig.~\ref{fig13}(b)) when a temperature dependent pure dephasing term of the form $\gamma_d$ = 1 +0.95(T-1) $\mu$eV~\cite{Ota,Borri} is added. Many of the finer features due to phonon modified spontaneous emission $\tilde\gamma$ are however lost.  The phonon modified SE rate $\tilde\gamma$ is around 2 $\mu$eV (T = 40 K) in the interesting regions outside the waveguide band, which is much smaller than $\gamma_d$. Thus the line-width of a QD is dominated by pure dephasing and do not change much as the QD frequency is scanned across the waveguide. Hence the intensity ratios are always stronger when the QD is to the right of a mode-edge, because phonon emission is stronger than absorption. 



\vspace{-0.01cm}


\section{Conclusions}\label{sec4}
In summary, we have provided an in-depth study of the incoherent emission spectra from QDs coupled to structured photonic reservoirs. We have carefully outlined the approximations made in different theories and applied them under suitable conditions to study emission spectra from photonic crystal cavities and waveguides. A non-Markovian approach is developed using a correlation expansion treatment for coupled QD-cavity systems,
and compared to the numerically much simpler approaches
of cQED and bath polaron master equation theories, and we discuss the pros and cons of these approaches for modelling QD cavity systems.
We also demonstrate the failure of the linear susceptibility approach for
explaining far off-resonant  cavity feeding. Using our general bath polaron master equation,
 a comprehensive study of the quantum emission spectra from weakly coupled 
 slow-light waveguides is provided where effects of the non-local LDOS on the SE rate are clearly seen. We highlight that our  general polaronic reservoir theory is completely general and can be applied to other structured reservoirs such as photonic band-edges, coupled cavities, finite-size waveguides and photonic molecules.

\appendix
\section{Coupled-mode spectrum from an inverted atom or electron-hole pair}
\label{Appen1}

\begin{figure}[t]
\vspace{0.cm}
\includegraphics[width=0.99\columnwidth]{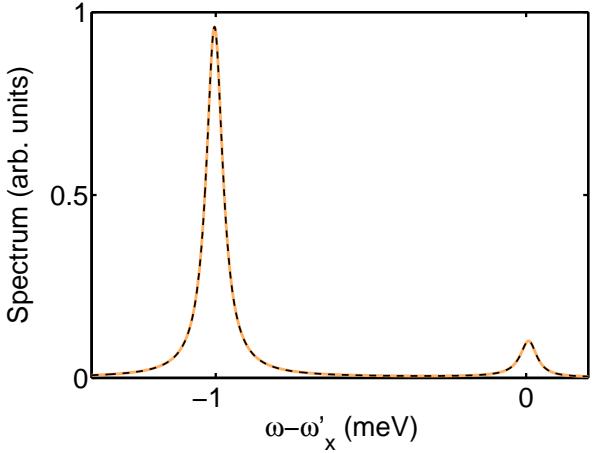}
\vspace{-0.cm}
        \caption{(Color online). \footnotesize{Normalized emission spectra at T = 4 K for a coupled QD-cavity system under off-resonant condition. The thick light solid (dark-dashed) line denote coupled-mode spectra $S^{\rm CM-cQED}_{\rm cav}$ ($S^{\rm CM-cQED, inv}_{\rm cav}$), calculated using polaron cavity-QED ME approach for an incoherenty excited atom (initally inverted QD). The main parameters are $g$ = 100 $\mu$eV, $\kappa$ = 65 $\mu$eV, $\gamma_0$ =  5 $\mu$eV, $\gamma_d$ =  55 $\mu$eV, $\Delta_{cx'}$ = 1 meV. }}
\label{fig15}
\end{figure}

In this appendix we compare the coupled mode spectrum calculated using the polaron cQED ME (\ref{eq9}), for a QD excited by a weak incoherent pump ($S^{\rm CM-cQED}_{\rm cav}$ solid light line, Fig.~\ref{fig15}) and that obtained from initially excited QD (i.e $\avg{\smdag\sm}(t=0) =1$) ($S^{\rm CM-cQED, inv}_{\rm cav}$ dark dashed line, Fig.~\ref{fig15}). The QD-cavity are off-resonant ($\Delta_{cx'} = 1$ meV and the spectras show perfect agreement as long as the QD is weakly excited in the first case ($S^{\rm CM-cQED}_{\rm cav}$). The reader should recall, that for an off-resonant high Q cavity (Fig.~\ref{fig6}(b)), the coupled mode spectrum $S^{\rm CM-cQED}_{\rm cav}$ derived using the polaron cQED ME is close to the physically correct coupled mode spectrum $S^{\rm CM-ce}_{\rm cav}$ calculated using correlation expansion approach. Thus the initial condition of an inverted atom correctly reproduces the spectra from a weakly driven QD. This clearly validates the initial condition used in linear susceptibility approach for calculation of linear spectra (Sec.~\ref{sec2d4}). 

When a QD is initially excited, the system does not have any steady-state excitation (cavity/QD) in absence of a drive and the coupled mode spectra is calculated by modifing (\ref{eq20})~\cite{Carmichael} as,

\begin{align}
\label{eqA1}
S^{\rm CM-cQED, inv}_{\rm cav}&(\omega) =
F(\r1_{\rm d},{\bf r}_{\rm D})\frac{\kappa}{\pi}\times \nonumber \\ &\text{Re}\left [\int_0^{\infty}dt \int_0^{\infty}d\tau\avg{a^{\dagger}(t+\tau)a(t)}e^{i(\omega_x'-\omega)\tau}\right]
\end{align}

For the calculations of Fig.~\ref{fig15} (light solid line) the weak incoherent pump produces a steady state QD excitation of $\avg{\smdag\sm}(t\rightarrow \infty) \approx 0.01$. When the pump is increased, the calculated spectra ($S^{\rm CM-cQED}_{\rm cav}$, light solid line) starts differing from the inverted atom spectra ($S^{\rm CM-cQED, inv}_{\rm cav}$, dark dashed line) when $\avg{\smdag\sm}(t\rightarrow \infty) > 0.1$ (not shown here). This corresponds to a $E/g >$ 0.01.

\acknowledgments
This work was supported by the Natural Sciences and Engineering Research Council of Canada and Queen's University.

\end{document}